\pdfoutput=1
\documentclass[aip,jcp,floatfix,reprint]{revtex4-1}

\usepackage{graphicx}
\usepackage{dcolumn}
\usepackage{bm}
\usepackage{color}
\usepackage{hyperref}

\newcommand{\hilightg}[1]{\colorbox{green}{#1}}
\newcommand{\hilighty}[1]{\colorbox{yellow}{#1}}

\begin{document}

\preprint{APS/123-QED}

\title{DNA hybridization kinetics: zippering, internal displacement and sequence dependence}
\author{Thomas E. Ouldridge,$^1$ Petr \v{S}ulc,$^1$ Flavio Romano,$^2$ Jonathan P. K. Doye$^2$ and Ard A. Louis }
\affiliation{$^1$Rudolf Peierls Centre for Theoretical Physics, Department of Physics, University of Oxford, 1 Keble
        Road, Oxford, UK, OX1 3NP \\
$^2$ Physical \& Theoretical Chemistry Laboratory, Department of Chemistry, University of Oxford,  South Parks Road, Oxford, UK, OX1 3QZ}

\begin{abstract}
While the thermodynamics of DNA hybridization is well understood, much less is known about the kinetics of this classic system.  Filling this gap in our understanding has new urgency because DNA nanotechnology often depends critically on binding rates. Here we use a coarse-grained model to explore the hybridization kinetics of DNA oligomers, finding that strand association proceeds through a complex set of intermediate states. Successful binding events start with the formation of a few metastable base-pairing interactions, followed by zippering of the remaining bonds. However, despite reasonably strong interstrand interactions, initial contacts frequently fail to lead to zippering because the typical configurations in which they form differ from typical states of similar enthalpy in the double-stranded equilibrium ensemble. Therefore, if the association process is analyzed on the base-pair (secondary structure) level, it shows non-Markovian behavior. Initial contacts must be stabilized by two or three base pairs before full zippering is likely, resulting in negative effective activation enthalpies. Non-Arrhenius behavior is observed as the number of base pairs in the effective transition state increases with temperature. In addition, we find that alternative pathways involving misbonds can increase association rates. For repetitive sequences, misaligned duplexes frequently rearrange to form fully paired duplexes by two distinct processes which we label `pseudoknot' and `inchworm' internal displacement. We show how the above processes can explain why experimentally observed association rates of GC-rich oligomers are higher than rates of AT-rich equivalents. More generally, we argue that association rates can be modulated by sequence choice.
\end{abstract}

\maketitle

DNA is central to biology, and has become a key ingredient in nanotechnology. Single strands of DNA have a sugar-phosphate backbone with bases (adenosine, thymine, guanine or cytosine -- hereafter referred to as A, T, G and C) attached at regular intervals. Watson and Crick\cite{Watson1953} showed that hydrogen bonding between A-T and G-C base pairs and stacking interactions between adjacent bases result in helical duplexes when two sequences  are complementary. This rule has been used to design structures,\cite{Pinheiro2011} machines\cite{Bath2007} and computational circuits\cite{Qian2011} that operate in parallel on a nanometer length scale. In many of these systems, assembly or operational dynamics is primarily driven by the association, or hybridization, of short strands of DNA (oligomers) to form duplexes of order ten base pairs. Understanding the details of oligomer association kinetics, and knowledge of how to accelerate or suppress reaction rates, is therefore essential if DNA nanotechnology is to fulfil its promise.

The thermodynamics of DNA duplex formation is well understood, and is dominated by states involving either strongly bound duplexes, or widely separated strands. Therefore, it is well characterized\cite{SantaLucia2004, Dirks2007} and can be described by all-or-nothing (two-state) models.\cite{SantaLucia2004}  By contrast, hybridization kinetics  depends on the rarely-visited intermediate states that lie between these two limits, and is therefore much harder to understand. In principle, the ensemble of transition pathways may be complex, leading to rich and subtle behaviour. Bimolecular association rate constants of $\sim 10^6 -10^7$\,M$^{-1}$s$^{-1}$ have been measured at approximately room temperature and at high salt concentrations ([Na$^+$] $\sim$ 1\,M or [Mg$^{2}$] $\sim$ 0.01\,M) for  DNA\cite{Gao2006, Morrison1993,Zhang_disp_2009} and RNA.\cite{Craig1971,Porschke1971,Porschke1973,Lima1992} There is agreement that dissociation rates increase exponentially with temperature \cite{Morrison1993,Craig1971,Porschke1971,Porschke1973,Chen_hybr_2007}, but authors have reported association rates that increase,\cite{Morrison1993, Porschke1973} decrease\cite{Craig1971,Porschke1971} and behave non-monotonically\cite{Chen_hybr_2007} with temperature. To our knowledge, there have been no systematic studies of the consequences of DNA sequence for hybridization rates.

Theoretical models at the level of secondary structure (the degree of base pairing) have been proposed to  explain experiments where association  rates decrease with temperature.\cite{Craig1971,Porschke1971,Chen_hybr_2007} These models posit that strands initially held together by short duplex sections tend to dissociate rather than fully hybridizing, either because they melt extremely quickly\cite{Craig1971,Porschke1971} or due to a thermodynamic barrier to full hybridization.\cite{Chen_hybr_2007} However, hairpins with stems as short as three base pairs are thermodynamically stable,\cite{Gao2006} and no detailed description of a barrier to completing hybridization has been proposed. 

Computer modelling can shed light on the details of hybridization kinetics. Ideally, simulations would use atomistic potentials such as AMBER\cite{Cornell1995} for maximal resolution, but the long time scales involved prevent exhaustive studies of such systems.  To explore reaction pathways, coarse-grained models are needed. These must be efficient enough to access the critical time-scales, but detailed enough to represent key features of the 3D structure, the mechanical properties  and thermodynamics of both single- {\em and} double-stranded DNA. A number of models have been proposed, but most are either `ladder' models that do not capture structural and mechanical properties of DNA\cite{Starr2006, Ouldridge2009, Araque2011,Linak2012, Svaneborg2012} or have not been carefully parameterized to DNA thermodynamics.\cite{Drukker2001,Morriss-Andrews2010,Dans2010,He2013} Hybridization kinetics have been studied in a detailed model known as 3SPN.1.\cite{Sambriski2008, Sambriski2009_PNAS} The authors identified initial binding and then `slithering' of strands past each other as a mechanism of duplex formation. However, single strands in 3SPN.1 are overly stiff and have structural and mechanical properties that are very similar to the duplex state, so association necessarily involves two pre-formed helices coming into contact. It is unclear whether the same pathway will be observed for a model with a more realistic description of the single-stranded state.


Here we apply a recently developed coarse-grained model, `oxDNA',\cite{Ouldridge2011,Ouldridge_thesis,Sulc2012} to the association of DNA oligomers.  The model incorporates the average structural, mechanical and thermodynamic properties of both single- and double-stranded DNA.  
Duplexes are stiff helices, but single strands can unstack, allowing them to adopt non-helical structures, reproducing their relative flexibility. Such flexibility allows oxDNA to capture properties such as the formation of hairpins\cite{Ouldridge2011} and the force-extension properties of single strands.\cite{Romano_overstretch_2012} We expect the flexibility of single strands to be a critical factor in hybridization. The robustness of oxDNA has been established by studying a range of phenomena that were not used for the initial parameterization. The formation of metastable kissing hairpins\cite{Romano2012}, cruciform structures under torsion\cite{Matek2012} and liquid crystals at high density\cite{DeMichele2012} are all reproduced in a physically reasonable way. Three dynamic DNA-based nanodevices have been simulated.\cite{Ouldridge_tweezers_2010, Sulc_walker_2012,Ouldridge_walker_2013} The calculations reproduced the designed behaviour, but also identified key subtleties arising from an interplay of structural, mechanical and thermodynamic factors.  OxDNA undergoes an overstretching transition, with the critical force in good agreement with experiment.\cite{Romano_overstretch_2012} Most importantly, oxDNA quantitatively reproduces\cite{Srinivas2013} the  
$10^{6.5}$-fold acceleration of the toehold-mediated strand displacement rate with increasing toehold length found by Zhang and Winfree\cite{Zhang_disp_2009}. As binding to the toehold involves the same self-assembly processes as in hybridization, our success gives us confidence to use oxDNA to study oligomer association  in detail.

We proceed as follows.  After briefly presenting the model and simulation techniques, we study hybridization processes for sequences designed to limit misbonding.   Hybridization involves a zipper-like mechanism, and reaction rates are suppressed at  increased temperature due to the instability of initial contacts. We then consider repetitive sequences, finding that  alternative pathways  to duplex formation, which we name `inchworm' and `pseudoknot' internal displacement, can significantly accelerate association.  Finally, we demonstrate that the instability of initial contacts and alternative pathways to assembly can lead to sequence-dependent hybridization rates, in agreement with some recent experiments.\cite{Zhang_disp_2009}

\section{Model and Methods}
\subsection{A coarse-grained model} 
\label{model}
OxDNA is detailed in Refs.~\onlinecite{Ouldridge2011}, ~\onlinecite{Ouldridge_thesis} and \onlinecite{Sulc2012}, and in Appendix \ref{app:model}. The version used for the majority of this work is given in Ref.~\onlinecite{Ouldridge_thesis}, and code implementing it is available for download.\cite{oxDNA}
A model strand is a chain of rigid bodies, each one representing a nucleotide. Nucleotides have one interaction site for the
backbone and two for the base. The potential energy of
the system includes terms for backbone connectivity, base-pairing, stacking and excluded volume interactions.


Base-pairing interactions are only included between complementary pairs A-T and G-C to reproduce Watson-Crick specificity. For much of this work, we use a parameterization with no further sequence dependence\cite{Ouldridge_thesis} to highlight generic properties that can be obscured by sequence-dependent effects. For sequence-dependent thermodynamics, we use a parameterization in which hydrogen-bonding and nearest-neighbour stacking strengths depend on the identity of the bases.\cite{Sulc2012} Both parameterizations were fitted to oligonucleotide melting temperatures predicted by SantaLucia's nearest-neighbour model,\cite{SantaLucia2004}
as well as the structural and mechanical properties of double- and single-stranded
DNA. The model was
fitted to experiments performed at  $[\mbox{Na}^{+}] =
0.5$\,M, a high salt concentration at which the strength of screening justifies incorporating electrostatic repulsion into a short-ranged excluded volume.

\subsection{Simulation techniques}
The majority of simulations in this work were performed using a  Langevin Dynamics (LD) algorithm\cite{Davidchack2009}. 
Langevin approaches represent an implicit solvent by augmenting the Newtonian equations of motion with drag and random noise forces. Simulated particles then undergo diffusive motion, and the whole system samples the canonical ensemble if the relative sizes of the drag and noise forces are chosen appropriately.\cite{Davidchack2009} Details of our implementation are given in Appendix \ref{methods:LD}. As is common in simulations of coarse-grained models, we use a higher diffusion coefficient than for physical DNA. The freedom to accelerate diffusion is an advantage of coarse-grained models, which also tend to exaggerate the speed with which processes occur by smoothing energy landscapes on a microscopic scale.\cite{Murtola2009} As a result, they can be used to study even more complex systems than would otherwise be expected. The price to pay is that only relative rates are physically meaningful, and we will focus on such relative rates in this work.


To check that the results reported here are not overly sensitive to the details of the simulation method and choice of friction constants, hybridization of non-repetitive duplexes was also simulated with diffusion coefficients reduced by a factor of 10. The results (Table \ref{simulations-high-friction}, discussed in Appendix \ref{methods:LD}) are qualitatively similar, except of course for an overall drop in the reaction rate with slower diffusion. For the sequence-dependent results at the end of this work, an alternative Brownian thermostat\cite{Russo2009} was used as detailed in Appendix \ref{methods:BD}. As we will show, the Brxownian algorithm (with a larger diffusion coefficient than the LD approach) produces behaviour consistent with the predictions of the LD thermostat, further evidence that the qualitative results of this work are not sensitive to the simulation method. 

To obtain good statistics for reaction transitions, which are dominated by rare events, we used Forward Flux Sampling (FFS).\cite{Allen2005,Allen2009} This technique facilitates sampling of a complex transition path ensemble by splitting a rare event into several stages that are easier to measure. Details of the application of FFS in this work are provided in Appendix \ref{methods:FFS}. Finally, simulations performed to measure equilibrium averages, rather than dynamics, were performed with an efficient cluster-move Monte Carlo algorithm,\cite{Whitelam2009} with the addition of umbrella sampling.\cite{Torrie1977} Details are given in Appendix \ref{methods:VMMC}.

\section{Results and Discussion}
\subsection{Hybridization of non-repetitive sequences}
\label{non-repetitive}
We first consider the hybridization of a 14-base duplex deliberately designed to limit non-intended base-pairing. The sequences of the two strands are:
\begin{itemize}
\item $5^\prime$-- TAT CTG GCT TGT CG -- $3^\prime$,
\item $5^\prime$-- CGA CAA GCC AGA TA -- $3^\prime$.
\end{itemize}
Simulations were run at a range of temperatures, from $300$\,K to $ 340.9$\,K, the latter being approximately the melting temperature of the strands at the concentration used. Details of the simulations are provided in Appendix \ref{hybr-nonrep}.
Additional simulations at 300\,K were performed in which only native (those expected in the full 14-base pair duplex)  base pairs were assigned a non-zero hydrogen-bonding energy, to determine the effect of non-native base pairs.


\begin{figure*}
\begin{center}
\includegraphics[width=18cm]{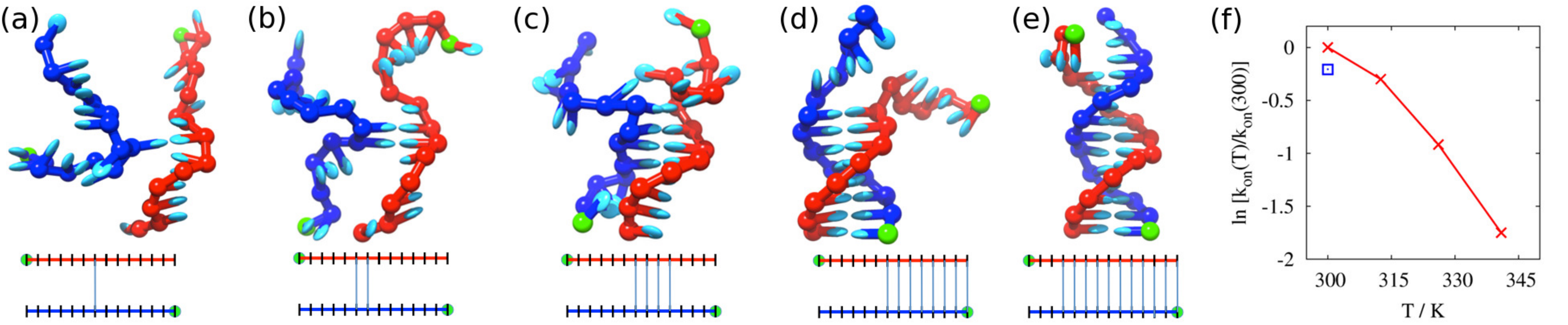}
\caption{(a)--(e) Stages of DNA hybridization, taken from a typical single trajectory. Green spheres indicate the $5^\prime$ end of each strand. Schematic diagrams underneath indicate the base pairs present in the system. (f) Hybridization rates $k_{\rm on}(T)$ of non-repetitive duplexes as a function of temperature $T$, relative to $k_{\rm on}({300})$, shown as red crosses connected by a solid line. Also shown (blue square) is the result for duplex formation  at $300$\,K when misaligned bonds are forbidden. \label{binding}}
\end{center}
\end{figure*}

Qualitatively, binding events start with the formation of a base pair between complementary bases after the strands have diffused into contact. In successful binding events, more base pairs subsequently form before this base pair breaks,  a process postulated elsewhere\cite{Craig1971,Porschke1971,Porschke1977,Saenger1984} and known as `zippering'. A typical pathway is illustrated in Fig. \ref{binding}\,(a)--(e). In some cases, initial base pairs are non-native, with native base pairs replacing them later. From Fig. \ref{binding}, it is clear that  zippering involves relatively unstructured single strands coming together to form base pairs in stages. Initial contacts can form between any bases  but have a bias towards those at the end of the strands, although initial contacts in the center are more likely to proceed to the full duplex once formed (see Fig. \ref{fig_flux_data}).

Fig. \ref{binding}\,(f) shows that  the binding rate decreases with increasing temperature. In the commonly used Arrhenius model of reaction kinetics, the association rate $k_{\rm on}$ depends on the temperature $T$ as
$k_{\rm on} = k_0 \exp (-H_{\rm a}/RT)$,
where $H_{\rm a}$ is a constant activation enthalpy, and $k_0$ is a constant rate. A system with a single well-defined transition state would indeed follow this prediction. A decrease in $k_{\rm on}$ with $T$ suggests a transition state with a negative enthalpy with respect to the unbound state. Overall, however, the Arrhenius model is a poor fit to our results (see Fig. \ref{fig_flux_data}\,(a)). The apparent activation enthalpy, which can be inferred from the slope $-\frac{{\rm d}}{{\rm d}(1/T)} \ln k_{\rm on}$, becomes more negative with temperature, with our model showing an apparent $H_{\rm a} (T)$ ranging from around $\sim -4.5$\,kcal\,mol$^{-1} $ to approximately $ - 12.5$\,kcal\,mol$^{-1}$  as $T$ rises from 300\,K to 340.9\,K. These values are similar to those measured for short RNA oligomers, which range from $ -5$\,kcal\,mol$^{-1}$\cite{Craig1971} to $ -9$ to $-18$ \,kcal\,mol$^{-1}$.\cite{Porschke1971} 

In contrast to association, we expect a relatively large positive activation enthalpy of dissociation because breaking a fully formed duplex involves disrupting many enthalpically favoured bonds (see  e.g. Fig. \ref{fig_FEP} ). This explains why experimental measurements find dissociation rates $k_{\rm off}$ that increase exponentially with increasing temperature.\cite{Morrison1993,Craig1971,Porschke1971,Porschke1973,Chen_hybr_2007}   Indeed, from calculations of the equilibrium constant $K_{\rm eq}=k_{\rm on}/k_{\rm off}$\cite{Ouldridge2011}, we find that $k_{\rm off}$  changes by about $\sim 10^{9}$ over the range 300--340.9\,K.   Here we focus on the more subtle behaviour of the association rate, which is especially relevant for non-equilibrium processes in DNA nanotechnology.

FFS allows us to sample from the ensemble of transition pathways.  At 300K, states involving two relatively well-formed
base pairs
have a 33\%
probability of reaching the full duplex,  while at 340.9K,  this success rate drops to just 8\% (Table \ref{results-non-repetitive}). Even for systems with only native base-pairing, states with two relatively well-formed base pairs still only progress to the full duplex in 65\% of cases at 300\,K. The fact that states with some base-pairing can fail to form a duplex explains the negative activation enthalpy: our effective `transition state' is enthalpically stabilized by base-pairing. Further, the typical number of base pairs in this `transition state' increases with temperature, as more base-pairing is required to make duplex formation probable. The reasons for the temperature dependence include: 1) the state with two base pairs itself becomes less stable, and 2) new bonds are less likely to form because a) strands become more unstructured and b) forming new base-pairs generates a smaller free-energy gain.   As a result, the activation enthalpy  becomes more negative with temperature  and there is no single  transition state with well-defined properties, explaining the non-Arrhenius behaviour. 

\begin{figure}
\begin{center}
\includegraphics[width=7cm]{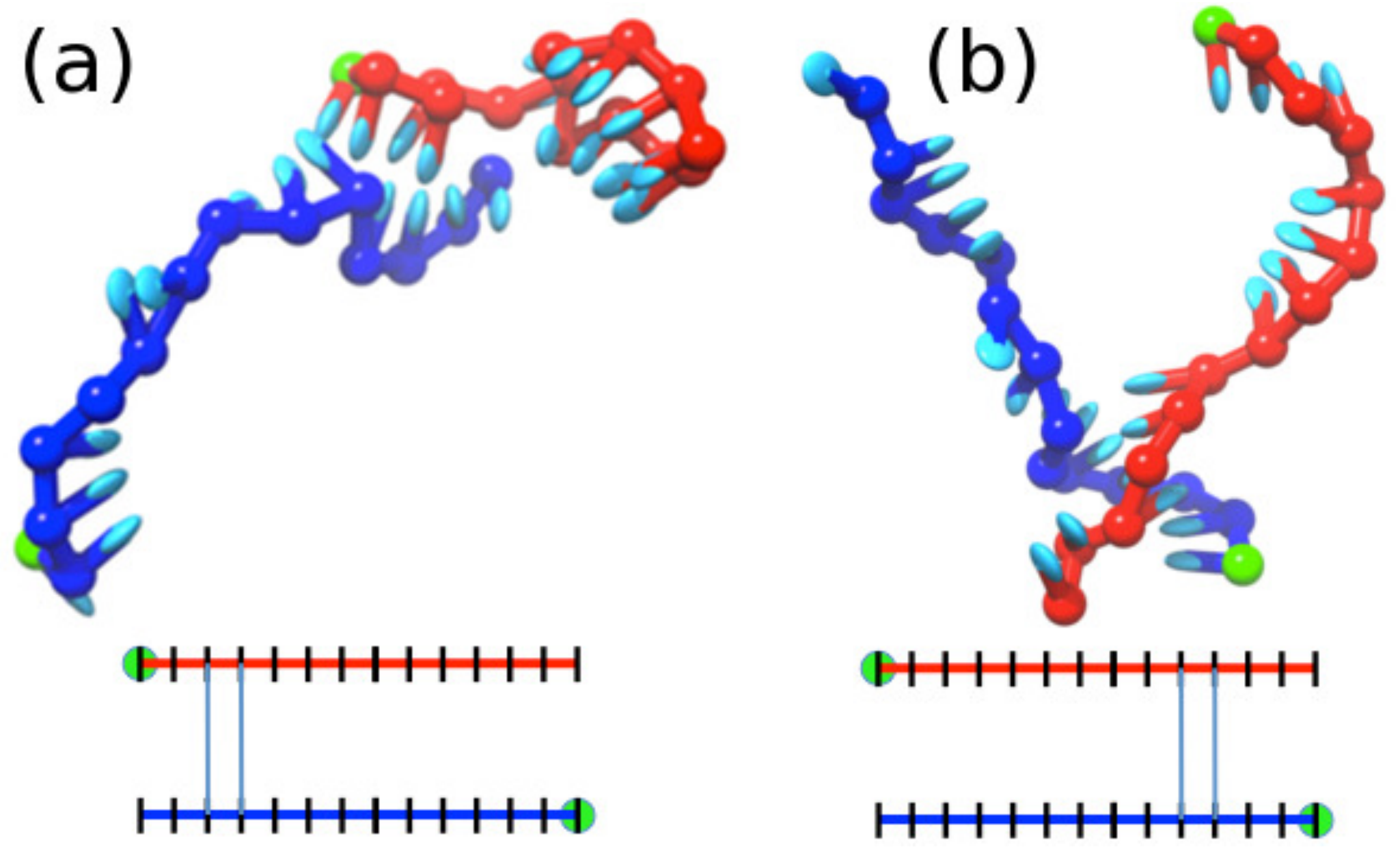}
\caption{Two configurations with two base pairs, and an interstrand enthalpy of $\sim -7.15$\,kcal\,mol$^{-1}$. The configuration in (a) is taken from a simulation of association, and that in  (b) from an equilibrium simulation of the system. \label{final-interface-image}}
\end{center}
\end{figure}

Simple thermodynamic considerations at the level of secondary structure do not explain  why many
initial contacts fail to completely hybridize. For example, free-energy profiles of the duplex states (see e.g. Fig. \ref{fig_FEP}) suggest that 
adding a single base pair reduces the free-energy of the system by 0.6\,kcal\,mol$^{-1}$ or $1.0 RT$ at 340.9\,K, and  1.7\,kcal\,mol$^{-1}$ or $2.9 RT$ at 300\,K. This argument suggests that the process should be favourable once the first base-pair has formed.   To understand why this reasoning fails we 
compare configurations obtained from hybridization simulations to configurations with
the same degree of base-pairing taken from equilibrium duplex
simulations. In Fig. \ref{final-interface-image}\,(a) and (b) we show two configurations with the same base-pairing
and overall interstrand enthalpy, with panel (a) obtained from a simulation of association initiated in the unbound state and panel (b) obtained from equilibrium simulations of the bound state (as detailed in Appendix \ref{protocols:eqm}).
Clearly the latter has a much more favourable 
spatial conformation for full duplex formation, since less rearrangement is required. A thorough analysis (Appendix \ref{results:eqm}) confirms that states with a certain number of base pairs found in assembly simulations are on average different from those in equilibrium simulations, and clearly  less conducive to full duplex formation. For example, the bases are further away from their native partners (see Table \ref{results-final-interface}).


The above argument requires that the breaking of the initial contacts can occur faster than strands equilibrate in the configuration space available given the existence of those contacts. This is plausible because the single strands are disordered. Thus the system appears non-Markovian when analysed only in terms of secondary structure: a given state has memory of whether it is accessed during assembly transitions or accessed from the bound ensemble. We stress that to observe such non-equilibrium effects, it is crucial to treat the three-dimensional  structure of single and double strands properly.

\subsection{Hybridization of repetitive sequences}
\label{repetitive}
Having studied strands in which non-native interactions are minimal, we now consider the limit  
of repetitive sequences that can form many misaligned structures (but no intrastrand hairpins). The sequences of the two strands are:
\begin{itemize}
\item $5^\prime$-- ACA CAC ACA CAC AC -- $3^\prime$,
\item $5^\prime$-- GTG TGT GTG TGT GT -- $3^\prime$.
\end{itemize}
At 300\,K we find a number of metastable structures in addition to the fully-bound duplex. These structures involve misaligned duplexes, which we label by their `register'. A register of $r$ corresponds to bases pairing with a partner offset by $r$ bases in the $5^\prime$ direction from their native partner. We find two basic classes, as outlined below.

\begin{figure*}
\begin{center}
\includegraphics[width=18cm]{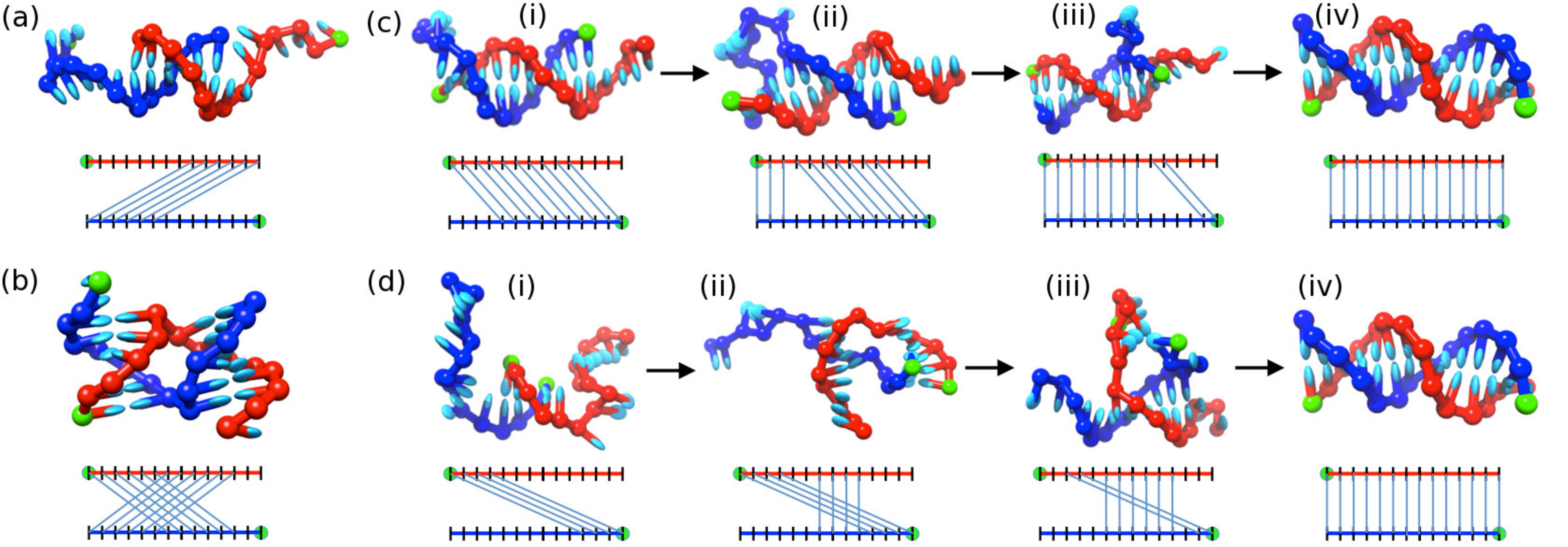}
\caption{(a) Misaligned bonding in register $-8$. (b) Pseudoknot bonding, in registers 6  and $-6$. Examples of internal displacement through  (c) the inchworm mechanism and (d) the pseudo knot mechanism. (c.i) The system is initially bound through 10 base pairs in register 4. (c.ii) Due to thermal fluctuations, three base pairs from register 0 form at the expense of three base pairs from register 4, resulting in a `bulge'. (c.iii) More base pairs in register 0 form at the expense of register 4. The bulge is passed down the helix. (c.iv) Eventually, the final base pair in register 4 breaks and register 0 is able to form all its possible base pairs. (d.i) A strand initially in register 10. (d.ii) Additional base pairs (from the correctly aligned register 0) form. (d.iii) Further base pairs form in register 0 at the expense of register 10. (d.iv) Eventually, the final base pair in register 10 breaks and a fully-bonded, correctly aligned duplex can form.}
\label{internal_disp}
\end{center}
\end{figure*}

\begin{itemize}
\item  Purely misaligned structures with the maximum number of base pairs given their register. A configuration with the maximal bonding for register $-8$ is illustrated in Fig. \ref{internal_disp}\,(a).
\item `Pseudoknot' structures,\cite{Dirks2007} characterised by two registers  $r_1$ and $r_2$. If we label nucleotides by their position on the strand (in the $5^{\prime} - 3 ^{\prime}$ direction) then in a pseudoknot the index of bases involved in pairing on one strand is a non-monotonic function of the index of their partner on the other strand. A typical metastable structure involving registers 6 and $-6$ is illustrated in Fig. \ref{internal_disp}\,(b).
\end{itemize} 
These metastable structures can be relatively slow to relax into either the fully-formed duplex or dissociated 
single strands. FFS is not efficient when intermediates with long lifetimes are present. We therefore initially measured the rates at  which strands formed a misaligned structure of at least four base pairs, or a number of the more stable pseudoknot states, at 300\,K. Further FFS simulations were performed to establish the eventual fate of a number of metastable states. Details are provided in Appendices \ref{hybr-rep}, \ref{disp-rep} and \ref{melt-rep}.


Strands can initially associate through zippering in a range of registers (Table \ref{results-repetitive}). The rate of formation of a given register is approximately proportional to the number of base pairs in that register, as a result of the number of possible initial contacts. From this point, systems with incomplete base pairing tend to rearrange into registers with a greater degree of base pairing. We describe these rearrangement processes as `internal displacement', as they involve the formation of a secondary double helix of an alternative register that competes for base-pairing with the first. This is analogous to the well known `strand displacement' process in which an invading strand replaces another within a duplex, except that in this case only two strands are involved (in a sense, a strand displaces itself). Two dominant rearrangement processes are observed. 
\begin{itemize}
\item `Inchworm' displacement (Fig. \ref{internal_disp}\,(c)): thermal fluctuations allow base pairs from an alternative register to form. The result is a `bulge' loop.\cite{SantaLucia2004} Generally, this (unfavorable) bulge is resolved by breaking the newly-formed base pairs in the alternative register. Occasionally, however, further base pairs  are broken in the  original register and additional base pairs in the new register form. The bulge can thus be passed through the original duplex in an inchworm fashion, allowing the new register to displace the old.
\item `Pseudoknot' displacement (Fig. \ref{internal_disp}\,(d)):  short misaligned duplexes have two long single-stranded tails. These tails can bind, resulting in a pseudoknot. The new register can compete for base pairs with the old, potentially displacing it. In some cases, such as the pseudoknot $-6,6$  in Fig. \ref{internal_disp}\,(b), neither arm of the pseudoknot can fully displace the other and some degree of spontaneous melting is necessary.  One of the arms in the pseudoknot can also be displaced (in an inchworm fashion) by an alternative register. 
\end{itemize}

Accurately measuring the transition rates between all registers is impractical due to the enormous number of possibilities and the large range of transition rates. Overall, however, initial alignments with more than 4 base pairs tend to undergo internal displacement to more strongly bound states, eventually reaching the full duplex. Misaligned duplexes of 4 base pairs frequently detach or undergo rearrangement. Internal displacement by a register with fewer base pairs than the original register is suppressed by the free-energy cost of breaking base pairs, although it is occasionally observed. 

At the low concentrations typical of experiment, the time spent in metastable intermediates is negligible compared to the typical time between attachment events. Metastable states then provide alternative pathways for the second-order process of association, increasing the rate constant for binding: in our case, by a factor of five for the repetitive sequences (see Appendix \ref{results-rep}). Repetitive sequences were also studied at 340.9\,K. At these temperatures,  short duplexes melt quickly and hence the probability that metastable structures are able to rearrange is reduced. Consequently the rate of formation of the fully-formed duplex falls off slightly faster with temperature than for the non-repetitive sequence: by a factor $\sim 9$ over the temperature range 300--340.9\,K rather than $\sim 5$. Internal displacement therefore provides another possible contribution to negative activation enthalpies in DNA duplex formation.   

\subsection{Sequence-dependence of binding rates}
\label{seq-dependence}
We have established two key facts: that initial contacts frequently dissociate before forming a full duplex, and that misaligned bonding can accelerate duplex formation through internal displacement. Both findings suggest possible mechanisms for sequence-dependence of DNA binding rates.
\begin{itemize}
\item In DNA, G-C base pairs are more stable than A-T. Hence, initial contacts between GC-rich sequences  should be more stable, and more likely to zip up following initial contact. If initial contacts form at approximately the same rate, duplexes with a greater density of G-C base pairs should form faster.
\item The number, stability with respect to dissociation and ease of internal displacement of misaligned metastable states will vary greatly from sequence to sequence. Increasing any of these factors should result in faster duplex formation.
\end{itemize}


Systematic experimental investigations of the sequence dependence of DNA binding rates are not available. Zhang and Winfree,\cite{Zhang_disp_2009} however, have probed sequence-dependent binding rates indirectly through toehold-mediated strand displacement. In the limit of long toeholds, the authors identified the displacement rate with the binding rate of toehold sequences. Their data showed a significant difference between the binding rates of GC-rich and AT-rich toeholds, with GC-rich toeholds causing faster displacement. To explore whether these differences can be attributed to the factors outlined here, we have simulated the association of 8-base duplexes using the sequences from Ref.  \onlinecite{Zhang_disp_2009}, as given in Table \ref{results-seq-dependent}. These simulations were performed using the Brownian thermostat with the sequence-dependent version of the model (more details are provided in Appendices \ref{methods:BD} and \ref{protocols:seq-dep}). The association rates (Table \ref{results-seq-dependent})  are in reasonable agreement with Ref. \onlinecite{Zhang_disp_2009}. The GC-rich sequence forms duplexes fastest, followed by the average-strength sequence and finally the AT-rich sequence. The GC-rich sequence is faster than the AT-rich analogue by a factor of 7.4: Zhang and Winfree found a factor of 15 in the long-toehold limit. Interestingly, part of the factor of 7.4 in our model can be attributed to correctly-aligned G-C base pairs being more stable than A-T equivalents, and part to an increased probability of binding via internal displacement. This conclusion follows from simulations in which internal displacement was suppressed by only allowing native base pairs: in this case, the GC-rich sequence was only 3.2 times as fast as the AT-rich variant.

\section{Conclusions}
 We have studied DNA hybridization using a coarse-grained model, oxDNA, that was carefully optimised to represent both single and double-stranded DNA. Stiff, helical duplexes form in a realistic fashion from flexible single strands.  By capturing these generic features of DNA we predict a complex ensemble of transition pathways for association, without a single transition state with well-defined properties, and qualitatively distinct dynamics for different sequences.  The association of a duplex  occurs through the formation of initial contacts involving a small number of bases, followed by zippering of the remainder, as suggested previously.\cite{Craig1971,Porschke1971,Porschke1977,Saenger1984} We go beyond this classic picture to show that initial contacts often dissociate, despite non-negligible attractive interactions, because their configurations are not conducive to full duplex formation and strands can detach before they equilibrate within the space of configurations defined by the secondary structure of the initial contacts.  Thus hybridization can fail even for interaction enthalpies which, if accessed from the equilibrium duplex ensemble, would overwhelmingly lead to duplex reformation.

Increasing the temperature destabilizes initial contacts, and lowers the drive to form more base pairs. The overall rate of association therefore decreases with temperature, resulting in a negative activation enthalpy if the results are interpreted through an Arrhenius model. At variance with the Arrhenius model, however, the effective activation enthalpy becomes more negative with increasing temperature, consistent with the fact that the strength of the initial contacts that are necessary to ensure duplex formation increases with temperature. Thus the system does not possess a single, well-defined `transition state' but a complex ensemble of transition pathways. This ensemble of pathways is further complicated by non-native interactions, which mean that systems can first form misaligned duplexes, and subsequently undergo internal displacement (rearranging without detaching) via {\it inchworm} or {\it pseudoknot} mechanisms to reach the fully formed duplex. As shown by the study of 8-base duplexes, sequences need not be perfectly repetitive for this pathway to be relevant.
At low reactant concentrations, these alternative pathways accelerate association. Due to the principle of detailed balance, dissociation must also occur via  internal displacement pathways, as well as via direct melting. We note that for longer strands, the probability of binding in a misaligned fashion is higher. We would therefore expect these mechanisms to contribute strongly to the association of longer strands. 

If initial contacts frequently fail, stronger contacts should prove more likely to succeed and thereby accelerate reaction rates. The rate of duplex formation through internal displacement will also depend strongly on the sequence. We tested the impact of these two proposed causes of sequence-dependent reaction rates and demonstrated their effect 
for short duplexes, finding agreement with experimental data. We predict that  systematic studies of hybridization rates for sequences  of varying GC content, and for sequences of equal overall binding strength but different degrees of repetitiveness will find that increasing GC content and repetitiveness accelerates duplex formation. We note that to resolve the effects clearly, care may have to be taken to avoid competing single-stranded hairpins.

It is worth contrasting our results with those found for  3SPN.1, an important model that has also been used to study hybridization.\cite{Sambriski2008,Sambriski2009_PNAS,Schmitt2011, Hoefert2011,Schmitt2013} These authors also observe complex kinetics, but with significant differences. In particular, for non-repetitive sequences, the authors claim that duplexes typically form non-native contacts, before `snapping' into the duplex state.\cite{Sambriski2009_PNAS} Other studies with the same model have found that the strands `wind' to form a double helix, then `slide' past each other to reach appropriate alignment.\cite{Schmitt2011,Schmitt2013} Repetitive sequences form misaligned structures, which relax into the fully-formed duplex by `slithering' past each other.\cite{Sambriski2008,Sambriski2009_PNAS} By contrast, the basic mechanism of duplex formation in oxDNA follows a clear nucleation and zippering pathway. Zippering occurs as bases from the relatively disordered single strands successively stack onto the growing duplex. Slithering, like internal displacement, allows misaligned duplexes to relax to the fully base-paired structure.\cite{Sambriski2009_PNAS} The mechanisms, however, are quite distinct: internal displacement involves the formation of two separate, base-paired duplex regions that compete for bases, whereas slithering involves the sliding of strands past each other in a process `devoid of significant energy barriers'.\cite{Hoefert2011} Both inchworm and pseudoknot mechanisms rely on the flexibility of the single strands and hence will be suppressed in 3SPN.1. In oxDNA, slithering is suppressed as it would require the system to pass through a double-helical state with no base-pairing. A more extensive explanation of the different mechanisms observed for 3SPN.1 and oxDNA is given in Appendix \ref{3SPN vs oxDNA}.

OxDNA is a simplified model, and it is therefore appropriate to evaluate the robustness of our conclusions. Firstly, the zipper-mechanism by which duplexes form relies on the existence of attractive interactions between bases, and the fact that the transition involves flexible single strands forming a stiff, helical duplex. These are generic features of DNA that are well reproduced by oxDNA, and hence the conclusion is likely to be reliable. Secondly, the internal displacement mechanisms identified involve  kinked and pseudoknotted  intermediates that are well-established motifs in nucleic acid secondary structure.\cite{SantaLucia2004,Pleij1985} Moreover, oxDNA describes the kinetics of conventional strand displacement involving three strands well.\cite{Srinivas2013} It is therefore likely that these pathways exist for real DNA. Thirdly, the frequent failure of initial contacts to form duplexes, despite the expected thermodynamic stability of extra base pairs, also relies on the well-established differences between single strands and duplexes. Nevertheless, it should be kept in mind that other contributions to the overall activation enthalpy may need to be taken into account. For example, we have not attempted to model the decrease of the viscosity of water with temperature, which may accelerate duplex formation at higher temperatures. Microscopic barriers such as the disruption of solvating water molecules prior to hydrogen-bond formation have also not been explicitly treated. In Appendix \ref{results:eqm} we show that initial contact between strands involves states with less intrastrand stacking on average than in the unbound ensemble. Breaking stacking helps the strands to be in contact  without being fully bound. This tendency contributes positively to the activation enthalpy -- in our model, this effect is smaller than the competing negative contributions, but the relative size may be different in nature.

Experimental studies do not currently provide a consistent picture of hybridization kinetics. In particular, it is not clear whether the rate of duplex formation typically increases or decreases with temperature (corresponding to positive or negative activation enthalpies respectively). We have, however, provided a physically reasonable justification for negative activation enthalpies; that initial contacts are surprisingly likely to detach because the overall configuration of the two strands is not conducive to full duplex formation. This argument explains why, in Markov models constructed at the base-pair level, one needs to either postulate a ``barrier'' to full duplex formation after the first few base pairs have formed,\cite{Chen_hybr_2007} or to make short sections of duplex detach very quickly.\cite{Craig1971,Porschke1971} Our simulations also suggest that if the failure of initial contacts is the cause of a negative activation enthalpy, precise experiments should show that this enthalpy becomes more negative with increasing temperature. Internal displacement can also enhance negative activation enthalpies. 

By providing insight into the complex mechanisms by which two DNA strands associate, we can thus suggest ways to modulate strand association rate and so choreograph the assembly and operation of DNA nanotechnology. Future work will consider the role of single-stranded hairpins in determining reaction kinetics, and the consequences of internal displacement for the association of longer strands.

\begin{acknowledgments}
The authors would like to thank E. Winfree, N. Srinivas, A. J. Turberfield and J. Bath for helpful discussions. T. E. O. acknowledges funding from Universtiy College, Oxford, and P.\v{S}. acknowledges support from the Scatcherd European Scholarship. This work was supported by the EPSRC.
\end{acknowledgments}

\appendix{}

\section{The DNA model}
\label{app:model}
\begin{figure}
\centering
\includegraphics[width=0.45\textwidth]{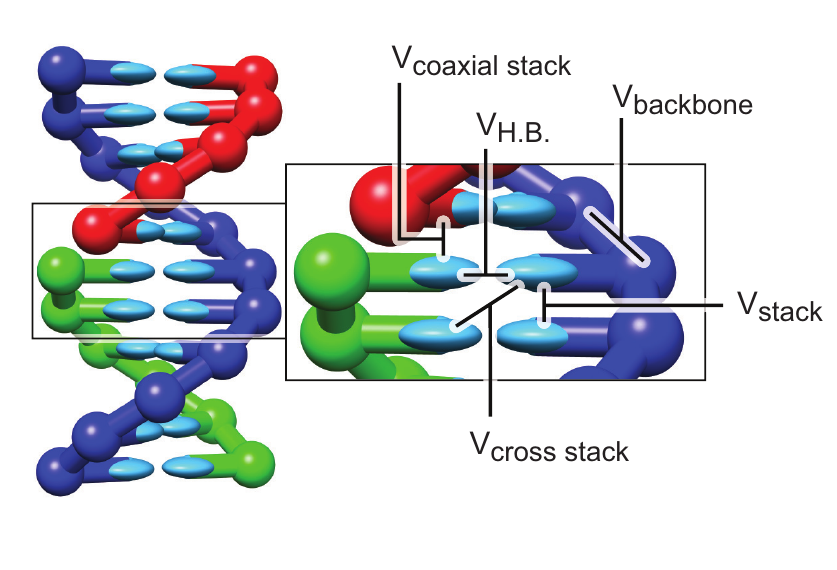}
\centering \caption{\footnotesize A model DNA duplex, with stabilising interactions depicted schematically. The backbone sites are shown as spheres, the bases as ellipsoids. Backbone colouring indicates strand identity. All nucleotides also interact with repulsive excluded volume interactions. The coaxial stacking interaction acts like a stacking interaction between bases that are not immediate neighbours along the backbone of a strand. Taken from Ref.\,\onlinecite{Ouldridge_walker_2013}.
\vspace{-0.1in}}
\label{fig_interactions}
\end{figure}

OxDNA and its interaction potentials have been described
in detail elsewhere.\cite{Ouldridge2011,Ouldridge_thesis,Sulc2012}
The model represents DNA as a string of nucleotides, where each nucleotide
(sugar, phosphate and base group) is a rigid body with interaction sites for
backbone, stacking and hydrogen-bonding interactions. The potential energy of
the system can be decomposed as 
\begin{eqnarray}
  V =  \sum_{\left\langle ij \right\rangle} \left( V_{\rm{b.b.}} + V_{\rm{stack}} +
V^{'}_{\rm{exc}} \right) + \nonumber \\
      \sum_{i,j \notin {\left\langle ij \right\rangle}} \left( V_{\rm HB} +  V_{\rm{cr.st.}}  +
V_{\rm{exc}}  + V_{\rm{cx.st.}} \right) ,
 \label{eq_hamiltonian}
\end{eqnarray}
where the first sum is taken over all nucleotides that are nearest
neighbors on the same strand and the second sum comprises all remaining
pairs. The interactions between nucleotides are schematically shown in 
Fig. \ref{fig_interactions}.  The backbone potential $ V_{\rm{b.b.}}$ is an isotropic
spring that imposes a finite maximum distance between backbone sites of neighbours, mimicking
the covalent bonds along the strand. The hydrogen bonding ($V_{\rm HB}$), cross
stacking ($V_{\rm{cr.st.}}$), coaxial stacking ($V_{\rm{cx.st.}}$) and
stacking interactions ($V_{\rm{stack}}$) are anisotropic and explicitly depend on
the relative orientations of the nucleotides as well as the distance
between the relevant interaction sites. This orientational dependence captures the planarity of bases, and helps drive the formation of helical duplexes. The coaxial stacking term is
designed to capture stacking interactions between bases that are not immediate neighbours along the backbone of a strand. Bases and backbones also have excluded volume interactions
$V_{\rm{exc}}$ or $V^{'}_{\rm{exc}}$.

Hydrogen-bonding interactions are only possible between complementary (A-T and C-G) basepairs. In the sequence-dependent parameterization, the strengths of interactions $V_{\rm{stack}}$ and $V_{\rm{HB}}$ further depend on the identity of the bases involved.\cite{Sulc2012} In the average model, $V_{\rm{stack}}$ is sequence-independent and $V_{\rm{HB}}$ is equivalent for  (A-T and C-G) base pairs. Interactions were  fitted to reproduce melting temperatures and transition widths of oligonucleotides, as predicted by SantaLucia's nearest-neighbor model.\cite{SantaLucia1998}
Note that our three dimensional model is significantly more complex than the nearest-neighbour model. We simply treat the latter as a high-quality fit to experimental data. Structural and mechanical properties of both double- and single-stranded DNA are also carefully taken into account in the fitting procedure. In DNA the double helical structure emerges because there is a length-scale mismatch between the preferred inter-base distance along the backbone, and the optimal separation of bases when stacking. It is exactly this feature that drives the helicity of oxDNA, rather than an imposed natural twist on the backbone. Overall, the emphasis in our derivation of oxDNA was on physics relevant for single-strand to duplex transitions. As discussed in the main text, oxDNA has been extensively tested for other DNA properties and systems to which it was not fitted. Our success in describing all these phenomena gives us confidence to use it to study the dynamics of hybridization.

OxDNA was fitted to reproduce DNA behavior at salt concentration $[\mbox{Na}^{+}] =
0.5$M, where the electrostatic properties are strongly screened, and it
is reasonable to incorporate them into a short-ranged excluded volume. The model therefore contains no further explicit electrostatic interactions. It should be noted that OxDNA neglects several features of DNA structure and interactions due
to the high level of coarse-graining. Specifically, the double helix in the
model is symmetrical rather than the grooves between the backbone sites having different sizes (i.e., major and minor grooving), and all four nucleotides have the
same structure. These differences with real DNA mean that oxDNA will not be able to treat phenomena that depend sensitively, for example, on anisotropic elasticity, explicit salt ion effects, or the existence of major and minor grooving. However, these specific properties of DNA are unlikely to be critical to the general arguments we are making about hybridization in this paper. Rather, it is the correct treatment of the basic mechanical properties of both single and double strands, together with the basic physics of hydrogen bonding and stacking that determines the emergent physical phenomena we are trying to describe.

 We assume that the partial pressure of DNA in dilute solution is negligible relative to that of our implicit solvent. In this limit, it is appropriate to consider the coarse-grained DNA strands in the canonical ensemble, even when comparing to experiments performed at constant pressure.\cite{Ouldridge_bulk_2012}  We therefore use the terms enthalpy and energy interchangeably, with enthalpy most natural for comparisons with the experimental literature, and energy most natural when discussing the implementation of the model.

\section{Simulation Methods}
\label{Simulation Methods}
The thermodynamic properties of  model DNA are given by averaging over the Boltzmann distribution
\begin{equation}
\rho({\bf r}^N, {\bf p}^N, {\bf q}^N, {\bf L}^N)  \hspace{2mm}  \propto \hspace{2mm} \exp({-\beta \mathcal{H}({\bf r}^N, {\bf p}^N, {\bf q}^N, {\bf L}^N) }).
\label{boltzmann distribution}
\end{equation}
In this equation $\mathcal{H}$ is the system Hamiltonian, ${\bf r}$ and ${\bf q}$ are positional and orientational coordinates and ${\bf p}$ and ${\bf L}$ are linear and angular momenta.
As terms in $\mathcal{H}$ containing ${\bf p}^N$ and ${\bf L}^N$ are separable and can be analytically integrated, the relative probability of a configuration is given by the Boltzmann factor  of its potential energy,  $\exp(-\beta V({\bf r}^N,{\bf q}^N))$.

Inferring model kinetics necessitates  an additional choice of dynamical algorithm. In this work we use Langevin Dynamics (LD) and a Brownian thermostat to measure  dynamical properties. Virtual Move Monte Carlo (VMMC) is also used to calculate thermodynamic averages. For the dynamical algorithms, it is necessary to define a nucleotide mass which is taken as $m=315.75$\,Da for all nucleotides.\cite{Ouldridge_thesis}  For dynamical purposes, we treat the nucleotides as spherical with a moment of inertia $31.586$\,Da\,nm$^2$.\cite{Ouldridge_thesis} The specification of mass, length and energy scales in the model together imply a time scale. For completeness, we will quote results  in the supplementary material terms of this time scale, although as discussed in the main text only relative times are physically meaningful.

\subsection{Langevin Dynamics}
\label{methods:LD}
LD is a formalism for including random and dissipative forces due to an implicit solvent in a self-consistent manner so that solute particles move diffusively and the system samples from the Boltzmann distribution. Newton's equations of motion for the solute particles can be augmented with these forces and integrated to give dynamical trajectories. The results reported in this work were obtained using the quaternion-based algorithm of Davidchack {\it et al.}\cite{Davidchack2009} To use LD, it is necessary to specify a friction tensor relating the drag forces experienced by a particle to its generalized momenta. We treat each nucleotide's interaction with the solvent as spherically symmetric, simplifying the friction tensors and leaving only two independent quantities, the linear and rotational damping coefficients $\gamma$ and $\Gamma$. We choose values of $\gamma = 0.59$\,ps$^{-1} $ and $\Gamma =1.76$\,ps$^{-1} $. These values produce overall diffusion coefficients of $D_{\rm sim} = 1.91 \times 10^{-9}$\,m$^{2}$s$^{-1}$ for a 14 base-pair duplex, higher than experimental measurements of $D_{\rm exp} = 1.19 \times 10^{-10}$\,m$^{2}$s$^{-1}$.{\cite{Lapham1997}} As discussed in the main text, accelerated diffusion is an advantageous aspect of coarse-grained modelling, allowing the simulations to access more complex processes. We show in Table \ref{simulations-high-friction} that using higher friction constants for the simple case of a non-repetitive sequence at 300\,K slows down hybridization, but does not qualitatively affect our results otherwise: in particular, the tendency for initial contacts not to proceed to full duplex  formation is preserved. LD Simulations in this work use a time step of 8.55\,fs. This time step has been previously shown to reproduce the energies and kinetics of shorter time steps for the DNA model.\cite{Ouldridge_thesis}

\subsubsection{Forward flux sampling}
\label{methods:FFS}
`Brute force' Langevin simulations are not always efficient enough to sample rare transitions. Forward flux sampling (FFS) allows the calculation of the flux between two local minima of free energy, and also samples from the trajectories that link the two minima ({\it reactive trajectories}).\cite{Allen2005,Allen2009}
Here we present a brief discussion of the FFS method in general. Our particular implementation will be discussed later.

The term `flux' from (meta)stable state $A$ to state $B$ has the following definition.
\begin{quotation}
Given an infinitely long simulation in which many transitions are observed, the flux of  trajectories from $A$ to $B$ is $\Phi_{AB} = N_{AB}/{(\tau f_{A})}$, where $N_{AB}$ is the number of times the simulation leaves $A$ and then reaches $B$,  ${\tau}$ is the total time simulated and $f_{A}$ is the fraction of the total time simulated for which state $A$ has been more recently visited than state $B$.
\end{quotation} 
The concept of  flux is therefore a generalization of a transition rate for processes that are not instantaneous: it incorporates the time spent in intermediate states  between $A$ and $B$. Subtleties relating to the inference of rates from our simulations are discussed in Appendix \ref{subtleties}.

To use FFS, we require an order parameter $Q$ which measures the extent of the reaction, such that non-intersecting interfaces, $\lambda^n_{n-1}$  can be drawn between consecutive values of $Q$. Initially, simulations are performed that begin in the lowest value of $Q$ (which we define as $Q=-2$), and the flux of trajectories crossing the surface $\lambda^0_{-1}$ (for the first time since leaving $Q=-2$) is measured. We define the lowest value of $Q$ as $Q=-2$ because the simulation procedure is distinct for $Q>0$.

The total flux of trajectories from $Q=-2$ to the alternative minima ($Q = Q_{\rm max}$) is then calculated as the flux across  $\lambda^0_{-1}$ from $Q=-2$, multiplied by the conditional probability that these trajectories reach $Q = Q_{\rm max}$ before returning to to $Q=-2$. This probability can be factorized into the product of the probabilities of trajectories starting from the interface $\lambda_{Q-1}^Q$ reaching the interface $\lambda_Q^{Q+1}$ before returning to $Q=-2$ to yield:
\begin{equation}
P(\lambda_{Q_{\rm max}-1}^{Q_{\rm max}} | \lambda_{-1}^0) = \prod_{Q=1}^{Q_{\rm max}}  P(\lambda_{Q-1}^Q | \lambda_{Q-2}^{Q-1}).
\label{FFS_eqn}
\end{equation}

\begin{figure}
\centering
\includegraphics[width=0.47\textwidth]{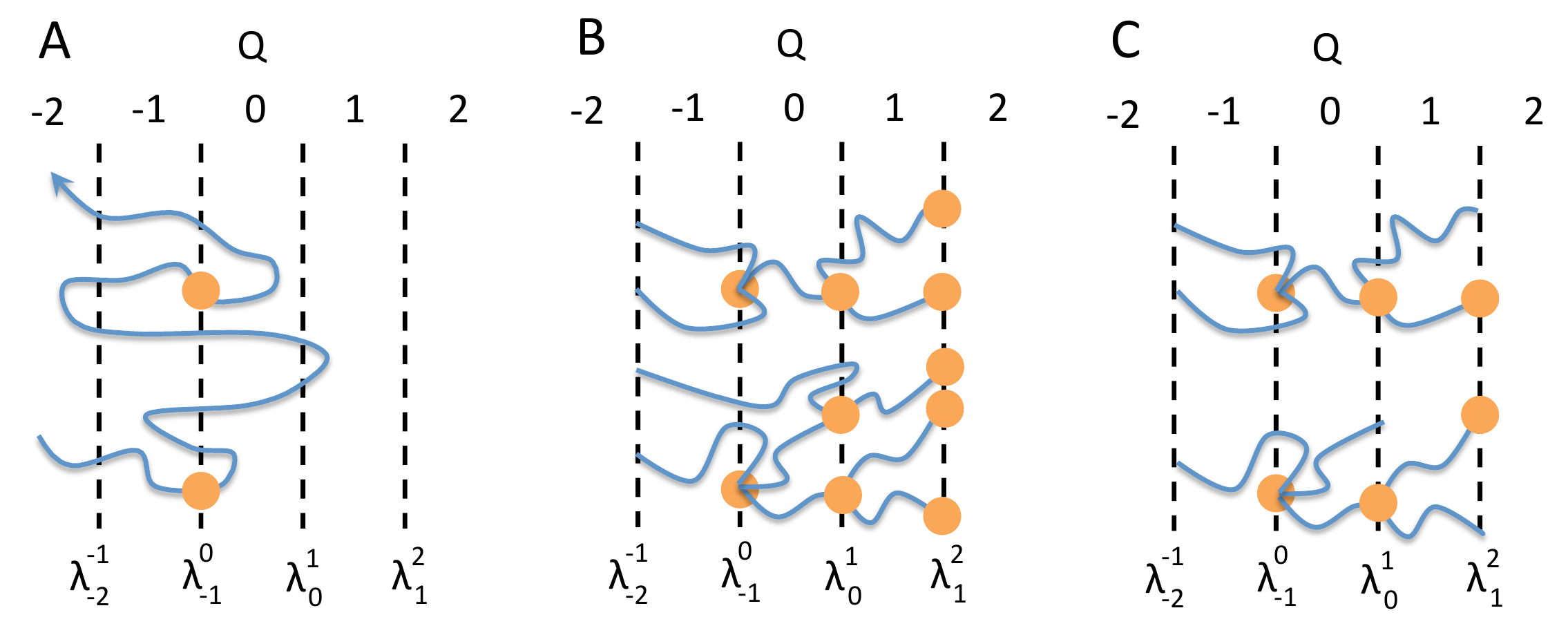}
\centering \caption{\footnotesize Schematic illustrations of FFS. An order parameter $Q$ is defined with interfaces $\lambda$ separating distinct value of $Q$. We are interested in measuring the flux from $Q=-2$ to $Q=2$ in this example. A) The initial measurement of flux across the interface $\lambda^0_{-1}$. Orange dots indicate the crossings that contribute to the flux, and also the states used to launch subsequent stages of simulation. B) Direct FFS involves randomly launching many trajectories from the interface $\lambda^0_{-1}$, and measuring the probability of reaching $\lambda^1_{0}$ before returning to $\lambda^{-1}_{-2}$. This procedure is then repeated for successive interfaces, resulting in branched trajectories. C) Rosenbluth FFS grows complete reactive trajectories in isolation: for each point on the previous interface, a fixed number of trajectories are launched. If one or more successfully reach the next interface, a single case is chosen at random and the rest are discarded. 
\label{fig_FFS}}
\end{figure}

In this work we use two distinct approaches to evaluating the product in equation \ref{FFS_eqn}, known as {\it direct} FFS and {\it Rosenbluth} FFS. Direct FFS proceeds by randomly loading microstates at the interface $\lambda^0_{-1}$ saved during the calculation of the initial flux, and using these as starting configurations from which to estimate $P(\lambda_{0}^1 | \lambda_{-1}^{0})$ by direct simulation. The process is then iterated for successive interfaces, using the successful trajectories from the previous interface as initial configurations for the next, allowing the estimation of $P(\lambda_{Q-1}^Q | \lambda_{Q-2}^{Q-1})$ for all relevant values of $Q$. Thus the flux from from $Q=-2$ to $Q=Q_{\rm max}$ can be calculated, and the trajectories obtained sample from the distribution of reactive trajectories. Rosenbluth FFS is an alternative approach in which, instead of successively performing a large number of simulations at every interface, individual reactive trajectories are generated independently by performing a small number of simulations at each interface for each trajectory. If multiple attempts are successful at a given interface, one is chosen at random and the rest are discarded. The contrasting approaches are illustrated in Fig. \ref{fig_FFS}. Note that extracting the flux and ensemble of reactive trajectories requires a re-weighting procedure for Rosenbluth FFS, due to the fact that some trajectories are discarded.\cite{Allen2005,Allen2009}

Direct FFS naturally produces branched trajectories, whereas Rosenbluth FFS does not. Rosenbluth sampling thus provides an equally good sampling of the initial stages of a reaction as the final stages, unlike direct FFS which samples the later stages in more detail. Unfortunately, however, Rosenbluth sampling is less efficient in obtaining reactive trajectories due to a tendency to generate successful simulations that are then discarded (to avoid branching). For these reasons we used Rosenbluth FFS where possible, but used direct FFS for the most difficult simulations.

The random error associated with FFS simulations can be estimated in the following way. Measurements of the flux across $\lambda_{-1}^0$ are performed using $N$ independent simulations. The daughter trajectories of any one of these initial simulations therefore give an independent estimate of the flux. We report the error as $\sigma /\sqrt{ N-1}$, where $\sigma^2$ is the variance of the $N$ independent estimates.

It is sensible to check that the FFS measurements are reasonable. This is possible because, during the sampling of the flux across interface $\lambda^0_{-1}$, successful transitions to $Q_{\rm max}$ are occasionally observed for the simplest cases. For example, during the simulation of the 14-base strands in which only native contacts were permitted, six transitions were observed in 72\,$\mu$s, giving a rate of $\sim 8\times 10^4$\,s$^{-1}$, consistent with the estimate of $(6.23\pm 0.47)\times 10^4$\,s$^{-1}$  from FFS (Table \ref{results-non-repetitive}).

\subsection{Brownian Thermostat}
\label{methods:BD}
A simple alternative to LD is to use an algorithm which evolves the system according to Newton's equations for a fixed length of time, then resample a fraction of the velocities and angular velocities from the Maxwell distribution.
We refer to this type of algorithm as a Brownain thermostat, and our simulations were performed using the thermostat described in Ref.~\onlinecite{Russo2009}. The simulation algorithm performs Verlet integration\cite{Frenkel2001}
for a given number of steps $N_{\rm Newt}$, then resets the velocity  of each nucleotide with probability $p_v$ and the angular velocity of each nucleotide with a probability $p_{\omega}$. The newly assigned velocities and angular velocities are drawn 
from the Boltzmann distribution. In our simulations, we chose  $p_v=0.02$, $p_\omega=0.0068$ and $N_{\rm Newt} = 103$. On time scales longer than $ N_{\rm Newt} \delta t / p_v$, where $\delta t$ is the integration time step, the dynamics is diffusive. Using $\delta t =8.53$\,fs, 14-base strands of DNA have a diffusion coefficient of $7.6 \times 10^{-8}$\,m$^2$s$^{-1}$ using this algorithm, higher than the $D_{\rm sim} = 1.91 \times 10^{-9}$\,m$^2$s$^{-1}$ measured for our LD algorithm.

\subsection{VMMC}
\label{methods:VMMC}
VMMC\cite{Whitelam2007,Whitelam2009} is a Monte Carlo technique effective for diluted systems with strong, directional interactions such as our DNA model. The  algorithm generates a series of configurations of a system that are drawn from the Boltzmann distribution. The algorithm moves from one configuration to the next by attempting moves of clusters that are generated in a manner that reflects local potential energy gradients in the system. Trial moves are accepted with a probability that ensures the system samples from the canonical ensemble. By moving clusters of strongly interacting particles, the algorithm is able to equilibrate model DNA systems much faster than simpler Monte Carlo algorithms. To use VMMC, it is necessary to select `seed' moves of a single particle: the resultant energy changes are used to generate the cluster. For all VMMC simulations reported here, the seed moves were:
\begin{itemize}
\item Rotation of a nucleotide about its backbone site, with the axis chosen from a uniform random distribution and the angle from a normal distribution with mean of zero and a standard deviation of 0.12 radians.
\item Translation of a nucleotide with the direction chosen from a uniform random distribution and the distance from a normal distribution with mean of zero  and a standard deviation of 1.02\,\AA. 
\end{itemize}

\subsubsection{Umbrella sampling}
\label{methods:umbrella}
Despite the simplicity of the model and the efficiency of VMMC, many processes are still slow to equilibrate due to the presence of large free-energy barriers. These barriers can be artificially flatenned, and equilibration enhanced, by incorporating an additional biasing weight 
$W({\bf r}^N, {\bf q}^N)$ \cite{Torrie1977}. In this approach, known as umbrella sampling, $W({\bf r}^N, {\bf q}^N)$ is chosen to favour the states of high free-energy, and the expectation of any variable $A$ can be extracted as
\begin{equation}
\langle A \rangle = \frac{\langle A({\bf r}^N, {\bf q}^N)/W({\bf r}^N, {\bf q}^N) \rangle_W }{\langle 1/W({\bf r}^N, {\bf q}^N) \rangle_W}.
\end{equation}
Here $\langle \rangle_W$ indicates sampling from the ensemble in which states have a relative probability \linebreak $W({\bf r}^N, {\bf q}^N) \exp(-\beta V({\bf r}^N, {\bf q}^N))$.

\subsection{Considerations of metastable states, fluxes and system size}
\label{subtleties}
\begin{figure}
\centering
\includegraphics[width=0.47\textwidth]{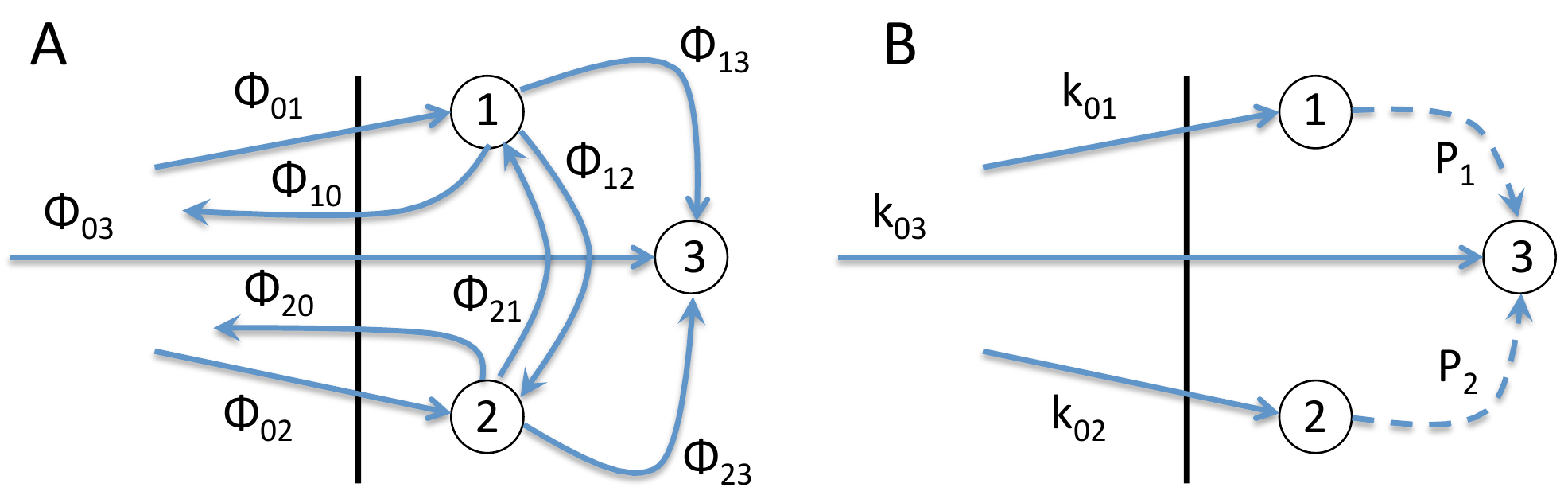}
\centering \caption{\footnotesize Reaction kinetics in the presence of metastable intermediates. A) Fluxes $\Phi$ contributing to the kinetics of formation of bound state ``3" from the unbound ensemble ``0" (taken as the left-hand-side of the diagram), in the presence of two metastable intermediates ``1" and ``2". B) Simplified picture used to interpret our results, in which each bound state is formed with a certain rate, and states 1 and 2 convert to the fully bound state 3 instantaneously with probabilities $P_1$ and $P_2$. The overall reaction rate is then $k_{03} + k_{01} P_1 + k_{02} P_2 $.
\label{fig_second_order_rates}}
\end{figure}

FFS is not effective at simulating transitions with long-lived metastable intermediates, as the process of escaping these intermediates must be directly simulated through brute-force methods. Such long lived intermediates are present with the repetitive sequences we have studied. FFS can, however, be used to simulate separately the flux of trajectories into and out of these states. This data can then be used to estimate overall reaction kinetics, by constructing a model such as that illustrated in \ref{fig_second_order_rates}\,A. Performing this calculation implicitly assumes that the system equilibrates within the metastable intermediate states before making another transition. 

In our work, we have employed this approximation to study repetitive sequences, defining a number of misbonded  intermediates and calculating the fluxes between them. The states considered are generally long-lived, show limited heterogeneity in structure within a state and are separated by significant free-energy barriers from other states. Thus the quasi-equilibrium assumption is reasonable.

Formation of either a metastable intermediate or the fully bonded state occurs in a short time scale after initial contact has been made, relative to the overall time spent in the single-stranded ensemble. Thus it is sensible
 to interpret the measured fluxes from the unbound ensemble directly as instantaneous reaction rates. Due to the existence of metastable intermediates for repetitive sequences, however, the {\it overall} process of moving from unbound to fully bound ensembles is not effectively instantaneous on the time scales that strands encounter one another through diffusion. In our simulations, a significant amount of time can be spent in misaligned registers  or pseudoknots. In principle, this makes the definition of an overall reaction rate problematic.

For computational tractability, however, we have simulated systems of two DNA strands with a relatively small volume, giving strand concentrations of $\sim 100$\,$\mu$M. This is much higher than typical experiments -- for instance, Zhang and Winfree\cite{Zhang_disp_2009} used a concentrations of order 1\,nM. In our system, all interactions are short ranged and hence the strands behave approximately ideally unless they come into close contact. Thus the effect of diluting the system is trivial: using a cell of twice the volume would halve the rate at which strands came into contact. Dilution should not, however, affect the rate at which strands dissociate or internally rearrange once attached, and nor will it affect the probability of choosing a given pathway out of a misbonded configuration. Therefore, although in our simulations a significant amount of time can be spent in metastable intermediates, at the much lower concentrations relevant to experiment these times will be negligible compared to the diffusional timescales required to make contact.

\begin{table*}[h]
\begin{center}
\begin{tabular}{c c c c c}
Order parameter  &  Separation  & Nearly-formed & Set of base pairs $A$ & Set of base pairs $B$ \\
$Q$ & $d/$nm & base pairs $n$  & with $E< E_{A}$ & with $E< E_{B}$  \\
\hline
\\
$Q=-2$ &  $d > 5.11$   &  $\sim$ & $\sim$ & $\sim$   \\
$Q=-1$ & $5.11 \geq d > 3.42$  & $\sim$ & $\sim$ & $\sim$ \\
$Q=0 $ &  $3.42 \geq d > 2.56$ &$\sim$ & $\sim$ & $\sim$\\
$Q=1 $ &  $2.56 \geq d > 1.71$  & $\sim$ & $\sim$ & $\sim$\\
$Q=2 $ &  $1.71 \geq d > 0.85$ & $\sim$ & $\sim$ & $\sim$ \\
$Q=3 $ & $ d \leq 0.85$  & $n = 0$ & $|A|=0$  & $|B|=0$\\
$Q=4 $ & $ \sim$ & $n \geq 1$ & $|A|=0$  & $|B|=0$ \\
$Q=5 $ & $ \sim$ & $\sim$ & \multicolumn{2}{c}{($|A| \geq 1$ \& $|B|=0$) or ($|A| =1$ \& $|B|=1$)} \\
$Q=6 $ & $ \sim$ & \multicolumn{3}{c} { $|A| \geq 2$ \& $|B| \geq 1$ \& ($A \notin S_r$ {\rm or} $B \notin S_r$ {\rm or} $n \neq 0^*$)} \\
$Q=7 $ & $ \sim$  & $n = 0^*$ & $A \in S_r$ &  $B \in S_r$ \\
\hline
\end{tabular}
\caption{\footnotesize Order parameter definitions for FFS simulations of binding. $|A|$ is the number of base pairs in set $A$.  $A \in S_r$ indicates that the set of interactions $A$ is in the set of sets $S_r$ that defines a target state of the simulations. For non-repetitive systems, only $S_0$ is relevant. For repetitive sequences at $T=300$\,K, 
misaligned structures with at least 4 base pairs ($14-|r_1| \geq 4$) are considered, and the pseudoknots $\{r_1,r_2\} = \{-4,8\}$, $\{-6,6\}$, $\{-6,8\}$, $\{-8,4\}$ and $\{-8,6\}$. For repetitive sequences at $T=340.9$\,K, only registers with at least eight base pairs are considered ($14-|r_1| \geq 8$), and pseudoknots are not. $n = 0^*$  indicates that nearly formed base pairs are forbidden, except in pseudoknot cases when nearly-formed base pairs from the relevant registers are allowed. The symbol `$ \sim$' indicates that no restriction is placed on this generalised coordinate except those that follow implicitly from the other requirements.  We use two energy cutoffs to monitor base-pairing: $E_A= -1.43$\,kcal\,mol$^{-1}$ and $E_B=-1.79$\,kcal\,mol$^{-1}$. \label{FFS-op-nonrep}}
\end{center}
\end{table*}

\begin{table*}[h]
\begin{center}
\begin{tabular}{c c c c c c}
& \multicolumn{5} {c} {Temperature/K}\\
 &  300 & 312.5 & 326.1 & 340.9 & $300^\prime$  \\
\hline 
\\
Number of simulations  & 20 &  20 & 20 & 20 & 20\\
for flux across $\lambda^0_{-1}$\\
\\
Initialization time  & 860& 860 & 860 & 860 & 860\\
per simulation /ns\\
\\
Crossings of $\lambda^0_{-1}$  &  8106 (75)  &8059  (69) &7948 (58) &8041 (50) & 7854 (72) \\
(time taken/$\mu$s)\\
\\
Flux across $\lambda^0_{-1}$ &$108 \pm 4$ &$116 \pm 2$ &$138 \pm 5$ & $160 \pm 8$& $109 \pm 4$\\
/$\mu$s$^{-1}$\\
\hline
\\
Total trajectories & 12000 & 4000 & 4000 & 4000 & 6000 \\ 
started from $\lambda^0_{-1}$\\
\\
Target interface &\multicolumn{5}{c}{Success probability (attempts per trajectory)} \\
$\lambda^1_{0}$ &  $0.427\pm 0.005$ (5)&  $0.420\pm 0.009$ (5)& $0.418\pm 0.008$ (5) & $0.427\pm0.008$ (5)&$0.431\pm 0.006$ (5)\\
$\lambda^2_{1}$ & $0.496\pm 0.003$ (5)&  $0.517\pm 0.006$ (5)&  $0.495\pm 0.007$ (5)& $0.514\pm0.008$ (5)& $0.504 \pm 0.004$ (5)\\
$\lambda^3_{2}$  & $0.521\pm 0.004$ (5)&  $0.528\pm 0.008$ (5)& $0.524\pm 0.006$ (5) & $0.545\pm0.006$ (5)& $0.514 \pm 0.006$ (5)\\
$\lambda^4_{3}$  & $0.534\pm 0.004$ (5)&  $0.567\pm 0.008$ (5)& $0.580\pm 0.009$ (5)& $0.609\pm0.008$ (5)& $0.275 \pm 0.006$ (5)\\
$\lambda^5_{4}$ & $0.146\pm 0.002$ (20)& $0.154\pm 0.004$ (40) & $0.165\pm 0.003$ (40)& $0.161\pm0.004$  (40)& $0.0685\pm 0.003$ (20)\\
$\lambda^6_{5}$  & $0.252\pm 0.01$ (5)& $0.182\pm 0.01$ (10)& $0.118\pm 0.008$ (10)& $0.0908\pm0.009$ (10)& $0.418 \pm 0.02$ (5)\\
$\lambda^7_{6}$  & $0.326\pm 0.02$ (5)& $0.265\pm 0.03$ (10)& $0.181\pm 0.02$ (10)& $0.0778\pm0.01$ (10)& $0.648\pm 0.02$ (5)\\
\hline
\\
Total reactive  & 1535 & 608 & 485 & 244 & 829 \\ 
trajectories found\\
\hline
\end{tabular}
\end{center}
\caption{Details of Rosenbluth FFS simulations performed to estimate the kinetics of binding for non-repetitive sequences at a range of temperatures ($300^\prime$ corresponds to simulations in which only native bonds had a non-zero interaction). The top half of the table reports the initial simulations of flux across the $\lambda^0_{-1}$ interface. The bottom half contains the simulation data from the later stages: the total number of trajectories initiated at $\lambda^0_{-1}$, the success rates of attempts to reach the next interface and the number of attempts per trajectory performed at each interface.\label{simulations-nonrep}}
\end{table*}

\begin{table}[h]
\begin{center}
\begin{tabular}{c c c}
 Friction coefficients & 0.59, 1.76 & 5.9, 17.6\\
 $\gamma$ /ps$^{-1} $, $\Gamma$ /ps$^{-1} $ &\\
 \\
 Difffusion coefficient & $1.91 \times 10^{-9}$ & $1.91 \times 10^{-10}$ \\
 / m$^2$s$^{-1}$ \\
\hline 
\\
Number of simulations & 20 & 20  \\
for flux across $\lambda^0_{-1}$\\
\\
Initialization time & 860 & 860\\
per simulation /ns\\
\\
Crossings of $\lambda^0_{-1}$ & 8106  (75) & 7734  (479) \\
(time taken/$\mu$s)\\
\\
Flux across $\lambda^0_{-1}$ & $108 \pm 4$ & $16.2 \pm 0.5$\\
/$\mu$s$^{-1}$\\
\hline
\\
Total trajectories & 12000 & 6000 \\ 
loaded from $\lambda^0_{-1}$ &\\
\\
Target interface & \multicolumn{2}{c}{Success probability} \\
&\multicolumn{2}{c}{(attempts per trajectory)} \\
$\lambda^1_{0}$ & $0.427\pm0.005$ (5) & $0.394\pm0.005$ (5)\\
$\lambda^2_{1}$ & $0.496\pm0.003$ (5)& $0.476\pm0.006$ (5)\\
$\lambda^3_{2}$ & $0.521\pm0.004$ (5)& $0.513\pm0.007$ (5)\\
$\lambda^4_{3}$& $0.534\pm0.004$ (5)& $0.630\pm0.005$ (5) \\
$\lambda^5_{4}$ & $0.146\pm0.002$ (20)& $0.258\pm0.003$ (20)\\
$\lambda^6_{5}$ & $0.252\pm0.01$ (5)& $0.190\pm0.008$ (5)\\
$\lambda^7_{6}$ & $0.326\pm0.02$ (5)& $0.267\pm0.02$ (5)\\
\hline
\\
Total reactive& 1535 & 584 \\ 
trajectories found\\
\hline
\end{tabular}
\end{center}
\caption{Comparison of Rosenbluth FFS simulations of hybridization for non-repetitive sequences at 300\,K for implementations of LD with different friction constants.  $\gamma = 0.59$\,ps$^{-1} $ and $\Gamma =1.76$\,ps$^{-1} $ are the standard values used  in this work.  The top half of the table reports the initial simulations of flux across the $\lambda^0_{-1}$ interface. The bottom half contains the simulation data from the later stages: the total number of trajectories initiated at $\lambda^0_{-1}$, the success rates of attempts to reach the next interface and the number of attempts per trajectory performed at each interface. \label{simulations-high-friction}}
\end{table}

\begin{table}[h]
\begin{center}
\begin{tabular}{c c c}
 & \multicolumn{2} {c} {Temperature/K}\\
 &  300 & 340.9 \\
\hline 
\\
Number of simulations & 50 &50  \\
for flux across $\lambda^0_{-1}$\\
\\
Initialization time & 860 & 860\\
per simulation /ns\\
\\
Crossings of $\lambda^0_{-1}$ & 19776  (186) & 19973  (133) \\
(time taken/$\mu$s)\\
\\
Flux across $\lambda^0_{-1}$ & $106 \pm 2$ & $150 \pm 3$\\
/$\mu$s$^{-1}$\\
\hline
\\
Total trajectories & 15000 & 6000 \\ 
loaded from $\lambda^0_{-1}$ &\\
\\
Target interface & \multicolumn{2}{c}{Success probability} \\
&\multicolumn{2}{c}{(attempts per trajectory)} \\
$\lambda^1_{0}$ & $0.434\pm0.004$ (5) & $0.417\pm0.004$ (5)\\
$\lambda^2_{1}$ & $0.505\pm0.003$ (5)& $0.504\pm0.005$ (5)\\
$\lambda^3_{2}$ & $0.519\pm0.003$ (5)& $0.532\pm0.005$ (5)\\
$\lambda^4_{3}$& $0.633\pm0.003$ (5)& $0.707\pm0.006$ (5) \\
$\lambda^5_{4}$ & $0.185\pm0.002$ (20)& $0.210\pm0.003$ (40)\\
$\lambda^6_{5}$ & $0.442\pm0.007$ (5)& $0.211\pm0.009$ (10)\\
$\lambda^7_{6}$ & $0.642\pm0.009$ (5)& $0.143\pm0.009$ (10)\\
\hline
\\
Total reactive& 5523 & 1207 \\ 
trajectories found\\
\hline
\end{tabular}
\end{center}
\caption{Details of Rosenbluth FFS simulations performed to estimate the kinetics of binding for repetitive sequences at 300\,K and 340.9\,K. The top half of the table reports the initial simulations of flux across the $\lambda^0_{-1}$ interface. The bottom half contains the simulation data from the later stages: the total number of trajectories initiated at $\lambda^0_{-1}$, the success rates of attempts to reach the next interface and the number of attempts per trajectory performed at each interface. \label{simulations-rep}}
\end{table}

 \begin{table*}[!]
\begin{center}
\begin{tabular}{c c c c c c}
Order parameter  &  Set of base pairs  &   Nearly-formed &Set of base pairs  &  Set of base pairs    & Set of base pairs   \\
$Q$ & $A$ &  base pairs $n$ & $B$  & $C$ not in $S_{r_0}$   &  $D$ not in $S_{r_0}$  \\
& with $E<E_A$ & & with $E<E_B$  &with $E<E_A$&with $E<E_B$ \\
\hline
\\
$Q=-2$ & $A \in S_{r_0}$ & $n=0$& $B \in S_{r_0}$ & $\sim$ & $\sim$   \\
$Q=-1$  & $\sim$ & \multicolumn{2}{c}{($B \in S_{r_0}$ \& $n\neq0$) or $B \notin S_{r_0}$ }& $|C|=0$ & $|D|=0$ \\
$Q=0 $ & $\sim$ & $\sim$ & $\sim$& \multicolumn{2}{c}{($|C|\geq1$ \& $|D|=0$) or  ($|C|=1$ \& $|D|=1$)} \\
$Q=1 $ & $\sim$ & $\sim$ & $\sim$&  \multicolumn{2}{c}{($|C| \geq 2$ \& $|D|=1$) or ($ |C| = 2$ \& $|D|=2$)}\\
$Q=2 $ & $\sim$ & $\sim$ & $\sim$ & \multicolumn{2}{c}{($|C| \geq 3$ \& $|D|=2$) or ($ |C| = 3$ \& $|D|=3$)}\\
$Q=3 $ &  \multicolumn{3}{c} {$A \notin S_{r^\prime}$ {\rm or} $n \neq 0^*$ {\rm or} $B \notin S_{r^\prime}$ } & $ |C| \geq 4$ & $|D| \geq 3 $  \\
$Q=4 $  & $A \in S_{r^\prime}$ & $n=0^*$ &  $B \in S_{r^\prime}$& $\sim$ & $\sim$\\
\\
\hline
\end{tabular}
\caption{\footnotesize Order parameter definitions for FFS simulations of internal displacement. $|A|$ is the number of base pairs in set $A$.  $A \in S_r$ indicates that the set of interactions $A$ is in the set of sets $S_r$. $S_{r_0}$ is the set corresponding to the initial (misaligned) state of the system. $S_{r^\prime}$ is the set corresponding to any {\it other} possible target state. For repetitive sequences at $T=300$\,K, 
misaligned structures with at least four base pairs ($14-|r_1| \geq 4$) are considered, and the pseudoknots $\{r_1,r_2\} = \{-4,8\}$, $\{-6,6\}$, $\{-6,8\}$, $\{-8,4\}$ and $\{-8,6\}$. For repetitive sequences at $T=340.9$\,K, only registers with at least eight base pairs ($14-|r_1| \geq 8$) are considered, and pseudoknots are not. $n = 0^*$  indicates that nearly formed base pairs are forbidden, except in pseudoknot cases when nearly-formed base pairs from the relevant registers are allowed. The symbol `$ \sim$' indicates that no restriction is placed on this generalised coordinate except those that follow implicitly from the other requirements. We use two energy cutoffs to monitor base-pairing: $E_A= -1.43$\,kcal\,mol$^{-1}$ and $E_B=-1.79$\,kcal\,mol$^{-1}$. During simulations, the system was also monitored to check for dissociation (when $d_{\rm min} > 5.11$\,nm). \label{FFS-op-rep-disp}}
\end{center}
\end{table*}

\begin{table*}[!]
\begin{center}
\begin{tabular}{c c c c c c}
{\bf A}  & \multicolumn{5} {c} {Initial register}\\
&  $2$ &  $4$ & $6$ & $8$ & $10$  \\
\hline 
\\
Number of simulations & 10 &  10 & 10 & 10 & 10\\
for flux across $\lambda^0_{-1}$\\
\\
Initialization time& 8.6 & 8.6 & 8.6 & 8.6 & 8.6\\
 per simulation /ns\\
\\
Crossings of $\lambda^0_{-1}$ &  1000 (39)  &1000  (54) &1000 (52) & 1000 (46)& 1000 (41)\\
(time taken/$\mu$s)\\
\\
Flux across $\lambda^0_{-1}$ &$25.9 \pm 0.6$ &$18.7 \pm 0.8$ &$19.4 \pm 0.7$ & $21.5 \pm 0.7$& $24.4 \pm 1$\\
/$\mu$s$^{-1}$\\
\hline
\\
Total trajectories  & 10000 & 5000 & 5000 & 6000 & 5000 \\ 
started from $\lambda^0_{-1}$ \\
\\
Target interface &\multicolumn{5}{c}{Success probability (attempts per trajectory)} \\
$\lambda^1_{0}$ &  $0.155 \pm0.006$ (20)& $0.258\pm0.01$ (20)& $0.275\pm 0.01$ (20)& $0.260\pm 0.01$ (10)& 0$.289 \pm 0.01$ (10)\\
$\lambda^2_{1}$ &  $0.239 \pm 0.02$ (20)& $0.232\pm0.02$ (20)& $0.224\pm0.01$ (20)& $0.261 \pm 0.02$ (10)& $0.401 \pm 0.02$ (3)\\
$\lambda^3_{2}$  &  $0.170 \pm 0.02$ (20)& $0.235\pm 0.02$ (20)& $0.248\pm0.01$ (20)& $0.504 \pm 0.02$ (10)& $0.860 \pm 0.02$ (3)\\
$\lambda^4_{3}$  &  $0.115 \pm 0.008$ (6)& $0.193\pm 0.008$ (6)& $0.355\pm0.03$ (6)& $0.807 \pm 0.03$ (10)& $0.971 \pm 0.005$ (2)\\
\hline
\\
Total reactive & 775 &  1052 & 1365 &  1309 & 1245\\ 
trajectories found \\
\hline
\\
\\
\bf {B} & \multicolumn{5} {c} {Initial register}\\
 &  $-2$ &  $-4$ & $-6$ & $-8$ & $-10$  \\
 \hline 
 \\
Number of simulations & 10 &  10 & 10 & 10 & 10\\
for flux across $\lambda^0_{-1}$\\
\\
Initialization time& 8.6 & 8.6& 8.6 & 8.6 & 8.6\\
 per simulation /ns\\
\\
Crossings of $\lambda^0_{-1}$ &  1000 (41)  &1000  (54) &1000 (54) & 1000 (50) & 1000 (42)\\
(time taken/$\mu$s)\\
\\
Flux across $\lambda^0_{-1}$ &$24.5 \pm 0.7$ &$18.4 \pm 0.6$ &$18.6 \pm 0.6$ & $20.2 \pm 0.6$& $24.1 \pm 0.5$\\
/$\mu$s$^{-1}$\\
\hline
\\
Total trajectories  & 10000 & 10000 & 10000 & 8500 & 8500 \\ 
started from $\lambda^0_{-1}$ \\
\\
Target interface &\multicolumn{5}{c}{Success probability (attempts per trajectory)} \\
$\lambda^1_{0}$ &  $0.182\pm0.009$ (20)& $0.266\pm0.01$ (20)& $0.273\pm0.02$ (20)& $0.276\pm0.01$ (10)& $0.308\pm0.009$ (10)\\
$\lambda^2_{1}$ & $0.240\pm 0.02$ (20)& $0.216\pm 0.02$ (20)& $0.243\pm0.02$ (20)& $0.282\pm0.01$ (10)& $0.436\pm0.02$ (3)\\
$\lambda^3_{2}$  & $0.221\pm0.03$(20)& $0.212\pm0.01$ (20)& $0.264\pm0.02$ (20)& $0.482\pm0.03$ (10)& $0.863\pm0.01$ (3)\\
$\lambda^4_{3}$  & $0.112\pm0.003$ (6)& $0.203\pm0.007$ (6)& $0.344\pm0.04$ (6)& $0.774\pm 0.02$ (10)& $0.960\pm 0.005$ (2)\\
\hline
\\
Total reactive & 823 & 2139 & 2731& 1957 & 2281  \\ 
trajectories found\\
\hline
\end{tabular}
\end{center}
\caption{Details of Rosenbluth FFS simulations performed to estimate the kinetics of internal displacement  at 300\,K, for positive (A) and negative (B) registers.  The top half of each table reports the initial simulations of flux across the $\lambda^0_{-1}$ interface. The bottom half contains the simulation data from the later stages: the total number of trajectories initiated at $\lambda^0_{-1}$, the success rates of attempts to reach the next interface and the number of attempts per trajectory performed at each interface. \label{results:internal_disp}}
\end{table*}

\begin{table*}[!]
\begin{center}
\begin{tabular}{c c c c}
  & \multicolumn{3}{c}{Initial register}\\
 &  $2$ & $4$ & $6$  \\
 \hline 
 \\
Number of simulations & 10 & 10  & 10\\
for flux across $\lambda^0_{-1}$\\
\\
Initialization time  & 8.6 & 8.6 & 8.6\\
per simulation /ns\\
\\
Crossings of $\lambda^0_{-1}$ &  1000 (11)  &1000  (11) &1000 (10) \\
(time taken /$\mu$s)\\
\\
Flux across $\lambda^0_{-1}$ & $92 \pm 2$ &$93\pm 3$ & $101 \pm 3$  \\
/$\mu$s$^{-1}$\\
\hline
\\
Total trajectories  & 6000 & 6000 & 6000\\ 
started from $\lambda^0_{-1}$ \\
\\
Target interface &\multicolumn{3}{c}{Success probability (attempts per trajectory)}\\
$\lambda^1_{0}$ & $0.195\pm0.008$ (20) & $0.248\pm0.01$ (20)& $0.234\pm0.01$ (20)\\
$\lambda^2_{1}$ & $0.294\pm0.02$ (20)& $0.330\pm0.01$ (20)& $0.313\pm0.02$ (20)\\
$\lambda^3_{2}$ & $0.284\pm0.02$ (20)& $0.323\pm0.01$ (20)& $0.433\pm0.03$ (20)\\
$\lambda^4_{3}$ & $0.142\pm0.01$ (6)& $0.223\pm0.01$ (6)& $0.311\pm0.02$ (6)\\
\hline
\\
Total reactive & 1285  & 921 &  971\\ 
trajectories found\\
\hline
\end{tabular}
\end{center}
\caption{Details of Rosenbluth FFS simulations performed to estimate the kinetics of internal displacement  at 340.9\,K. The top half of the table reports the initial simulations of flux across the $\lambda^0_{-1}$ interface. The bottom half contains the simulation data from the later stages: the total number of trajectories initiated at $\lambda^0_{-1}$, the success rates of attempts to reach the next interface and the number of attempts per trajectory performed at each interface. \label{internal_disp_highT}}
\end{table*}

\begin{table*}[!]
\begin{center}
\begin{tabular}{c c c c c c c}
\hline
&   \\
Order parameter  &  Separation   &  Set of base   &   Nearly-formed &Set of base    & Number of base & Number of base  \\
$Q$ &$d/$nm  & pairs $A$ &  base pairs $n$ & pairs $B$  & pairs $X$ from $S_{r_0}$   & pairs $Y$ from $S_{r_0}$ \\
& & with $E< E_A$ & &   with $E< E_B$ &   with $E< 0$ &   with $E< E_Y$ \\
\hline \hline
\\
$Q=-2$ & $\sim$ & $A \in S_{r_0}$ & $n=0$& $B \in S_{r_0}$ & $\sim$ & $\sim$  \\
$Q=-1$  & $\sim$ &   \multicolumn{3}{c} {$A \notin S_{r_0}$ {\rm or} $n \neq 0$ {\rm or} $B \notin S_{r_0}$ } &  $ X \geq 13 - |r_0|$ & $Y \geq 13 - |r_0| $   \\
$Q=0$ & $\sim$ & $\sim$ & $\sim$ & $\sim$ & \multicolumn{2}{c}{($ X < 13 - |r_0|$ \& $ Y < 13 - |r_0|$)}\\
& & & & & \multicolumn{2}{c}{\&} \\
& && & & \multicolumn{2}{c}{($ X > 12 - |r_0|$ {\rm or} $ Y > 11 - |r_0|$)}\\
$Q=m$, $1 \leq m <12-|r_{0}| $ & $\sim$ &$\sim$ & $\sim$& $\sim$ & \multicolumn{2}{c}{($ X \leq 13- |r_{0}| -m $ \& $ Y \leq 12-|r_0| -m$)}  \\
& & & & & \multicolumn{2}{c}{\&} \\
& & & & &  \multicolumn{2}{c}{($ X >12- |r_{0}| -m$ or $ Y > 11-|r_{0}| -m$)}  \\
$Q=12-|r_0| $ & $d <5.11$ &  \multicolumn{3}{c} {$A \notin S_{r^\prime}$ {\rm or} $n \neq 0^*$ {\rm or} $B \notin S_{r^\prime}$ } & $X \leq1$ & $Y=0$ \\
$Q=13-|r_0| $ & $d \geq 5.11$  & $\sim$& $\sim$& $\sim$& $\sim$& $\sim$\\
\\
\hline
\end{tabular}
\caption{\footnotesize Order parameter definitions for direct FFS simulations of dissociation of misaligned duplexes at 300\,K. Order parameters were different for different initial states $r_0$ due to the differing number of base pairs. $14-|r_0|$ gives the total number of base pairs possible in the initial misaligned register $r_0$. $r^\prime$ is any {\it other} possible bound structure. misaligned structures with more than four base pairs are considered as possible structures $r^\prime$, along with the pseudoknots $\{r_1,r_2\} = \{-4,8\}$, $\{-6,6\}$, $\{-6,8\}$, $\{-8,4\}$ and $\{-8,6\}$. $n = 0^*$  indicates that nearly formed base pairs are forbidden, except in pseudoknot cases when nearly-formed base pairs from the relevant registers are allowed. The symbol `$ \sim$' indicates that no restriction is placed on this generalised coordinate except those that follow implicitly from the other requirements. During simulations, the system was also monitored to check for the formation of alternative bound structures, through the criteria $A \in S_{r^\prime}$ \& $n=0^*$ \&  $B \in S_{r^\prime}$. The energy scales  $E_A= -1.43$\,kcal\,mol$^{-1}$, $E_B=-1.79$\,kcal\,mol$^{-1}$ and $E_Y= -0.596$\,kcal\,mol$^{-1}$ are used here.
\label{FFS-op-melting}}
\end{center}
\end{table*}

\begin{table*}[!]
\begin{center}
\begin{tabular}{c c c c c c c}
\hline
&   \\
Order parameter  &  Separation   &  Set of base   &   Nearly-formed &Set of base    & Number of base & Number of base  \\
$Q$ &$d/$nm  & pairs $A$ &  base pairs $n$ & pairs $B$  & pairs $X$ from $S_{r_0}$   & pairs $Y$ from $S_{r_0}$ \\
& & with $E< E_A$ & &   with $E< E_B$ &   with $E<0$ &   with $E< E_Y$ \\
\hline \hline
\\
$Q=-2$ & $\sim$ & $A \in S_{r_0}$ & $n=0$& $B \in S_{r_0}$ & $\sim$ & $\sim$  \\
$Q=-1$  & $\sim$ &  \multicolumn{3}{c} {$A \notin S_{r_0}$ {\rm or} $n \neq 0$ {\rm or} $B \notin S_{r_0}$ } &  $ X \geq 13 - |r_0|$ & $Y \geq 13 - |r_0| $   \\
$Q=0$ & $\sim$ & $\sim$ & $\sim$ & $\sim$ & \multicolumn{2}{c}{($ X < 13 - |r_0|$ \& $ Y < 13 - |r_0|$)}\\
& & & & & \multicolumn{2}{c}{\&} \\
& && & & \multicolumn{2}{c}{($ X > 11 - |r_0|$ {\rm or} $ Y > 10 - |r_0|$)}\\
$Q=m$, $1 \leq m <6-\frac{1}{2}|r_{0}| $ & $\sim$ &$\sim$ & $\sim$& $\sim$ & \multicolumn{2}{c}{($ X \leq 13- |r_{0}| -2m $ \& $ Y \leq 12-|r_0| -2m$)}  \\
& & & & & \multicolumn{2}{c}{\&} \\
& & & & &  \multicolumn{2}{c}{($ X >11- |r_{0}| -2m$ or $ Y > 10-|r_{0}| -2m$)}  \\
$Q=6-\frac{1}{2}|r_{0}| $ & $d <5.11$ &  \multicolumn{3}{c} {$A \notin S_{r^\prime}$ {\rm or} $n \neq 0$ {\rm or} $B \notin S_{r^\prime}$ } & $X \leq1$ & $Y=0$ \\
$Q=7-\frac{1}{2}|r_{0}|$ & $d \geq 5.11$  & $\sim$& $\sim$& $\sim$& $\sim$& $\sim$\\
\\
\hline
\end{tabular}
\caption{\footnotesize Order parameter definitions for direct FFS simulations of melting at 340.9\,K. Order parameters were different for different initial configurations, due to the differing number of base pairs. $14-|r_0|$ gives the total number of base pairs possible in a misaligned register $r_0$, with $r_0$ being the initial register of the system. $r^\prime$ is any {\it other} possible bound structure. misaligned structures with at least eight base pairs are considered as possible structures $r^\prime$, and pseudoknots are not included.The symbol `$ \sim$' indicates that no restriction is placed on this generalised coordinate except those that follow implicitly from the other requirements. During simulations, the system was also monitored to check for the formation of alternative bound structures, through the criteria $A \in S_{r^\prime}$ \& $n=0$ \&  $B \in S_{r^\prime}$.
\label{FFS-op-melt-highT}}
\end{center}
\end{table*}

\begin{table*}[!]
\begin{center}
\begin{tabular}{c c c c c}
 & \multicolumn{4} {c} {Initial register}\\
 &  $10$ &  $-10$ & $8$ & $-8$  \\
\hline 
\\
Simulations run & 2 & 8 & 10 & 9\\
for flux across $\lambda^0_{-1}$\\
\\
Initialization time  & 8.5 &8.5 & 8.5 &8.5\\
per simulation /ns\\
\\
Crossings of $\lambda^0_{-1}$ &  39453  (5.3)  &79915  (9.0) & 99442 (9.4) &90096 (8.4)\\
(time taken /$\mu$s)\\
\\
Flux across $\lambda^0_{-1}$ & $7.33^*$  &$8.78 \pm 0.28$ & $10.6 \pm 0.1$  & $10.6 \pm 0.1$\\
/ns$^{-1}$\\
\hline
\\
Target interface & \multicolumn{4}{c}{Total attempts/successes at later stages}\\
$\lambda^1_{0}$ & 20000 / 856 & 50000 / 2107 & 130000 / 6745 & 140000 / 7509\\
$\lambda^2_{1}$ &40000 / 1497& 90000 / 1807 & 150000 / 10253 & 150000 / 10016\\
$\lambda^3_{2}$  &3000 / 374& 10000 / 2285& 50000 / 3484 & 50000 / 3437\\
$\lambda^4_{3}$  &N/A & N/A & 20000 / 3689  & 20000 / 4336\\
$\lambda^5_{4}$  & N/A & N/A & 12488 / 230 & 7000 / 111\\
\hline
\end{tabular}
\end{center}
\caption{Simulation results for direct FFS simulations of melting of misaligned duplexes at 300.0\,K. The top half of the table describes the  initial flux simulations, and the bottom half contains the data from subsequent interfaces. The number of reactive pathways in direct FFS is simply the number of successes at the final interface. 
 $^*$Only two initial calculations of flux were run for this state, making the calculation of errors unreliable. Note that, particularly for the registers $\pm 10$, internal displacement processes were also observed during simulations. These processes are not well described by the FFS order parameter of melting, meaning that they are not accurately sampled in these simulations. An uneven presence of these displacement trajectories for registers $\pm 10$  causes the large differences between individual entries for the two registers in the table. The overall rate of melting (Table \ref{results-repetitive_2}), however, is similar for the two cases, as it should be. This also suggests that the  error on the register 10 estimate is reasonably small.  \label{FFS-melt-data}}
\end{table*}

\begin{table*}[!]
\begin{center}
\begin{tabular}{c c c c}
& \multicolumn{3} {c} {Initial register}\\
&  $6$ & $4$ & $2$   \\
\hline
\\
Simulations run & 15 & 48 & 20\\
for flux across $\lambda^0_{-1}$\\
\\
Initialization time & 8.5 & 8.5 & 8.5 \\
 per simulation /ns\\
 \\
Crossings of $\lambda^0_{-1}$ &  149747 (7.7)  &479021 (23) &401195 (18) \\
(time taken /$\mu$s)\\
 \\
Flux across $\lambda^0_{-1}$  &$19.4 \pm 0.4$ & $20.3 \pm 0.2$  & $22.2 \pm 0.2$\\
/ns$^{-1}$\\
\hline 
\\
Target interface & \multicolumn{3}{c}{Total attempts/successes at later stages}\\
\\
$\lambda^1_{0}$& 160000 / 11789 & 260000 / 18607 & 250000 / 18150\\
$\lambda^2_{1}$ & 23000 / 10425 & 245000 / 10791 & 250000 / 11634\\
$\lambda^3_{2}$ & 122500/12577 & 33750 / 5174 & 32783 / 4795 \\
$\lambda^4_{3}$ & 5000 / 2717 & 4500 / 587  & 24000 / 3963\\
$\lambda^5_{4}$ & N/A & 10000 / 3417 & 20500 / 4618\\
$\lambda^6_{5}$  & N/A &  N/A & 9000 / 1399\\
\hline
\end{tabular}
\end{center}
\caption{Simulation results for direct FFS simulations of melting of misaligned duplexes at 340.9\,K. The top half of the table shows how many independent initial flux simulations were performed, and the initialization time for each of these simulations before data was recorded. The bottom half contains the simulation data: the total crossings of $\lambda^0_{-1}$ and the time simulated in the first stage, and the total attempts and successes of the later stages. The number of reactive pathways in direct FFS is simply the number of successes at the final interface. \label{FFS-melt-highT-data}}
\end{table*}

When we compare relative rates to experiment, the time required to form an initial contact should scale with the dilution, and thus {\it relative} rates of initial attachment measured in simulation are directly comparable to experiment. Any time spent in metastable intermediates, however, should be subtracted from the total time to reach a duplex from the single-stranded state in order to make a fair comparison.
Practically, this means that when analysing the repetitive sequence, we simply use the measured fluxes out of each metastable state to calculate the probability that a given intermediate will progress to the fully formed duplex before dissociating. The overall reaction rate $k_{\rm on}$ is then 
\begin{equation}
k_{\rm on} = \sum_i k_i P_i,
\end{equation}
where the sum runs over all bound states $i$, $k_i$ is the formation rate of $i$ from the unbound ensemble and $P_i$ is the probability that such a state will convert to the fully-formed structure before dissociating. We illustrate this analysis schematically in Fig.\,\ref{fig_second_order_rates}\,B.  For the sequence-dependent study, we simply subtract the time in which the two strands have one or more base pairs from the total time that it takes to reach the fully-bound state. We thus compensate for the fact that rearrangement is a significant contribution to reaction times at our concentrations, but not in the dilute limit.

\section{Simulation protocols}
\label{protocols}
In this section we discuss the implementation of the algorithms of Appendix \ref{Simulation Methods} for the specific systems studied in this work, and present some raw data from the simulations which would allow reproduction of the results.  Processed data is presented in Appendix \ref{processed_data}. To facilitate the discussion, we introduce the following concepts.

\begin{itemize}
\item {\em Native} base pairs are those that are expected to form in the fully-bound structure. 
\item For repetitive sequences, a number of metastable intermediates exist. Some of these are simply misaligned structures, which can be unambiguously defined by their {\em register}: a register of $r_1$ corresponds to bases pairing with a partner offset by $r_1$ bases in the $5^\prime$ direction from their native partner. For non-repetitive sequences, only $r_1=0$ is relevant. The maximum number of base pairs in a register $r_1$ is $14- |r_1|$.
\item Other structures involve two registers $\{r_1,r_2\}$ in a pseudoknotted configuration. We found that   $\{r_1,r_2\} = \{-4,8\}$, $\{-6,6\}$, $\{-6,8\}$, $\{-8,4\}$ and $\{-8,6\}$ were long-lived metastable states, as displacement of one arm by the other is limited.
\item  Given criteria for identifying base-pairing interactions, a given state of the system will have a set of interactions $A$. Let  $A$ be in the set of sets $S_{r}$ if  $A$ is characteristic of the bonding pattern in register $r = r_1$ or pseudoknot $r=\{r_1,r_2\}$. For purely misaligned structures, we take $A$ to be in $S_{r_1}$ if every possible base pair in register $r_1$ is present, with no other interactions. Pseudoknots have a greater degree of heterogeneity (base pairs can be exchanged between registers). For our purposes, a set of interactions $A$ is in $S_{r_1,r_2}$ if and only if each of the registers has at least 6 interactions, and no other interactions are present.
\item The separation $d_{\rm min}$ is the minimum distance between hydrogen-bonding sites over all pairs of bases in the two strands. 
\item One way of identifying interactions is through the {\em nearly formed} base pair. A potential base pair between the strands is counted as {\em nearly formed} when the conditions outlined below hold.
\begin{itemize}
\item The separation of hydrogen-bonding sites is $\leq 0.85$\,nm.
\item The hydrogen-bonding potential consists of a separation dependent factor multiplied by a number of modulating angular factors. At most one of these factors that contributes multiplicatively to the hydrogen-bonding energy is zero.  
\item The hydrogen-bonding interaction is less negative (weaker) than $-1.43$\,kcal\,mol$^{-1}$. Typical hydrogen bonds have enthalpies of  $\sim -3.6$\,kcal\,mol$^{-1}$.
\end{itemize}
 Physically, these conditions mean that the bases are close and fairly well aligned, but not forming a strong base pair. 
\end{itemize}

\subsection{Hybridization of non-repetitive duplexes}
\label{hybr-nonrep}
Studies of hybridization were performed using Rosenbluth FFS. The order parameter $Q$, as detailed in Table \ref{FFS-op-nonrep}, combines separation-based and interaction strength metrics. Its complicated form is designed to optimize sampling, and reduce the possibility of two interfaces being crossed in a single integration time step (this is aided by the use of two different energy cutoffs for quantifying the degree of base-pairing).  The condition for $Q=7$ indicates a bound state in this work, and will be relevant in other cases. For the repetitive sequences, we simultaneously measure the flux into a number of possible binding registers. Data from the simulations are given in Table \ref{simulations-nonrep}.

\subsection{Hybridization of repetitive duplexes}
\label{hybr-rep}
Measurements of the initial stage of attachment were performed exactly analogously to the non-repetitive duplexes, and the order parameter is outlined in Table \ref{FFS-op-nonrep}. In this case, the flux of trajectories into several metastable structures was measured simultaneously. At 300\,K, purely misaligned structures with at least four base pairs  ($r_1 = 0,\pm 2, \pm 4, \pm 6, \pm 8, \pm 10$) were considered, as well as the pseudoknots $\{r_1,r_2\} = \{-4,8\}$, $\{-6,6\}$, $\{-6,8\}$, $\{-8,4\}$ and $\{-8,6\}$. At 340.9\,K, only  misaligned structures with at least eight base pairs were considered as metastable targets, because other structures melt rapidly and require no explicit treatment. Details of the simulations are given in Table \ref{simulations-rep}.

\begin{table*}[t]
\begin{center}
\begin{tabular}{c c c c c c c c c c c c c c c c}
\hline
Number of base pairs & 0 & 1 & 2 & 3 & 4 & 5 & 6 & 7 & 8 & 9 & 10 & 11 & 12 & 13 & 14 \\
\hline 
\\
Biasing weight   & 0 & $ 10^{14} $& $5\times 10^{12}$ &$2\times 10^{11}$ & $ 10^{10}$ & $5 \times 10^8$ & $2\times10^7$ & $10^6$ & $5 \times 10^4$ & 2000 & 100 & 5 & 0.2 & 0.01 & $5\times 10^{-4}$ \\
 $W$ at 300\,K \\
Biasing weight  &  0 & $3 \times 10^5$ &$10^5$ &$3 \times 10^4$ &$10^4$ &$3000$ &$1000$ &$300$ &$100$ & 30 &10 & 3 & 1 & 1 & 1\\
$W$ at 340.9\,K \\
\hline
\end{tabular}
\end{center}
\caption{Umbrella potential used to bias equilibrium simulations of the duplex state. A base pair counts as formed if it has an energy $E < -0.596$\,kcal\,mol$^{-1}$. \label{umbrella bias}}
\end{table*}

\subsection{Internal displacement of misaligned repetitive sequences} 
\label{disp-rep}
Measurements of internal displacement were performed using Rosenbluth FFS. The order parameter is outlined in Table \ref{FFS-op-rep-disp}.  At 300\,K, we consider the flux of trajectories from any misaligned structure to any other alignment with at least 4 base pairs, as well as the metastable pseudoknot states $\{r_1,r_2\} = \{-4,8\}$, $\{-6,6\}$, $\{-6,8\}$, $\{-8,4\}$ and $\{-8,6\}$. The relaxation of these metastable pseudoknots proved to be too difficult to simulate reliably. At 340.9\,K, only alignments with at least eight base pairs were considered. Simulations were monitored to check for strand dissociation: trajectories that resulted in dissociation were ended and counted as `failures' for the purposes of measuring the flux of internal displacement. Further details of the simulations are given in Table \ref{results:internal_disp}.

\subsection{Dissociation of misaligned repetitive sequences} 
\label{melt-rep}
Measurements of dissociation of misaligned structures were performed using direct FFS. The order parameter is outlined in Tables \ref{FFS-op-melting} and \ref{FFS-op-melt-highT}. At 300\,K, dissociation was studied for registers $r_1 = \pm 8$ and $\pm 10$. Longer misaligned structures have a dissociation rate that is negligible with respect to internal rearrangement. Simulations were monitored to check for rearrangement into alternative metastable states: trajectories that resulted in internal displacement were ended and counted as `failures' for the purposes of measuring the dissociation flux. 

At 340.9\,K, where dissociation is much faster, it was studied for registers $r_1 =2$, 4 and $6$. Registers $-2$, $-4$ and $-6$ should be related to $2$, 4 and $6$ by the symmetry of the model, and were not simulated for the sake of efficiency. The other results presented here show no significant asymmetry between positive and negative registers, justifying this approach. Further details of the simulations are given in Tables \ref{FFS-melt-data} and \ref{FFS-melt-highT-data}.

\subsection{Characterization of the equilibrium ensemble}
\label{protocols:eqm}
To understand the kinetic results, it is helpful to characterize the equilibrium ensemble of duplex states. VMMC simulations were performed on a pre-formed 14-base-pair duplex at 300\,K and 340.9\,K, with umbrella sampling used to enhance the sampling of states with a low degree of base-pairing (but forbid full detachment). The bias applied is detailed in Table \ref{umbrella bias}. Four simulations were run for $10^9$ VMMC steps at each temperature, with an initialization period of $10^6$ steps. For simplicity, simulations were performed on systems in which only native base-pairing was permitted. During these simulations, the properties of states with base-pairing energies consistent with the penultimate FFS interface of association were saved.  The possible states are given below. Following the notation of Table \ref{FFS-op-nonrep}, let  $E_A= -1.43$\,kcal\,mol$^{-1}$ and $E_B=-1.79$\,kcal\,mol$^{-1}$.  The two classes of states are:
\begin{enumerate}
\item One base pair with $E< E_B$ and one  other base pair with $E \approx E_A$, with no other base pairs with $E_B<E< E_A$.
\item One or more base pairs with $E_B<E< E_A$ and one other base pair with $E \approx E_B$, with no other base pairs with $E<E_B$.
\end{enumerate} 
In practice, to sample these states, $E \approx E_X$ was taken as $E = E_X \pm 0.03$\,kcal\,mol$^{-1}$. The equilibrium probability $P(n)$ of $n$ base pairs with an energy of $E<-0.596$\,kcal\,mol$^{-1}$ being present was also measured.

We also compared the intrastrand enthalpy (primarily arising from nearest-neighbor stacking interactions) in the single-stranded ensemble at 300\,K to the value obtained from averaging over configurations at the penultimate FFS interface of association.  The single-stranded ensemble was sampled in an identical fashion to the duplex simulations above, except that the umbrella potential was set to unity in the absence of base pairs, and zero if any base pairs were present. All states were used to calculate the average intrastrand enthalpy.

\subsection{Sequence-dependence of association rate}
\label{protocols:seq-dep}
Pairs of 8-base strands were simulated using the Brownian thermostat in a periodic cell of volume $2.09 \times 10^{-21}$\,l, at a temperature 298.15\,K (the temperature used in the experiment of Zhang and Winfree\cite{Zhang_disp_2009}).  For each sequence, the time taken for association into the full duplex structure was measured 1000 times. In each case, the system was initialized in the same single-stranded configuration, but with distinct nucleotide velocities. Any correlation resulting from using the same configuration is minimal, as the shortest association time in the simulations is three orders of magnitude larger than the equilibration and diffusion time scales in the single-stranded state.  As discussed in Appendix \ref{subtleties}, the time spent in  structures with interactions present between the two strands was not included in this estimate of the association time. Errors in the estimates of rates were calculated using the standard error on the mean of 20 independent estimates, each obtained from 50 events. Additional simulations were performed in which base pairing interactions were restricted to native contacts. 

\section{Results}
\label{processed_data}

\subsection{Hybridization of non-repetitive sequences}
\begin{table}[!]
\begin{center}
\begin{tabular}{c  c  c c }
Allowed base  & $T/{\rm K}$ & flux / s$^{-1}$ & 2\,bp success \\
pairs& & & probability \\
\hline
\\
Any & 300.0 & $(7.67\pm 0.75) \times 10^{4}$ & $0.33 \pm 0.018$ \\
Any & 312.5 & $(5.68 \pm 0.77) \times 10^{4}$ & $0.27 \pm 0.028$ \\
Any  & 326.1 & $(3.06\pm 0.43) \times 10^{4}$ & $0.18 \pm 0.016$\\
Any & 340.9 & $(1.34\pm 0.24) \times 10^{4}$ & $0.078\pm 0.010 $\\
Native only & 300 & $(6.23\pm 0.47)\times 10^4$ & $0.65 \pm 0.023$\\
\hline
\end{tabular}
\caption{\footnotesize Total flux from unbound to fully bound state for non-repetitive sequences. Also shown is the probability of successful completion of duplex formation once the system has reached the penultimate FFS interface. \label{results-non-repetitive}}
\end{center}
\end{table}

\begin{figure*}[t]
\begin{center}
\includegraphics[angle=-90, width=0.3\textwidth]{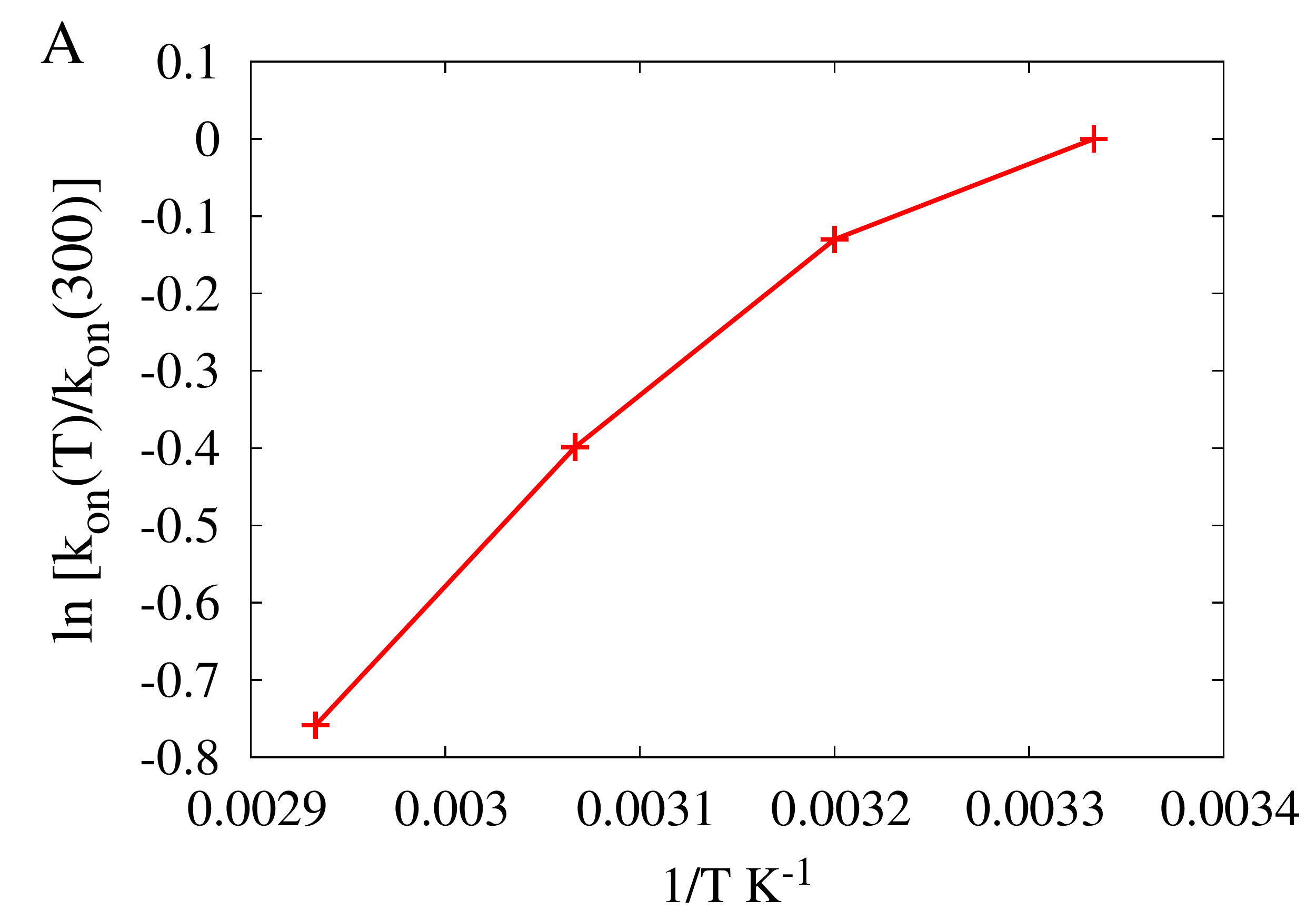}
\includegraphics[angle=-90, width=0.3\textwidth]{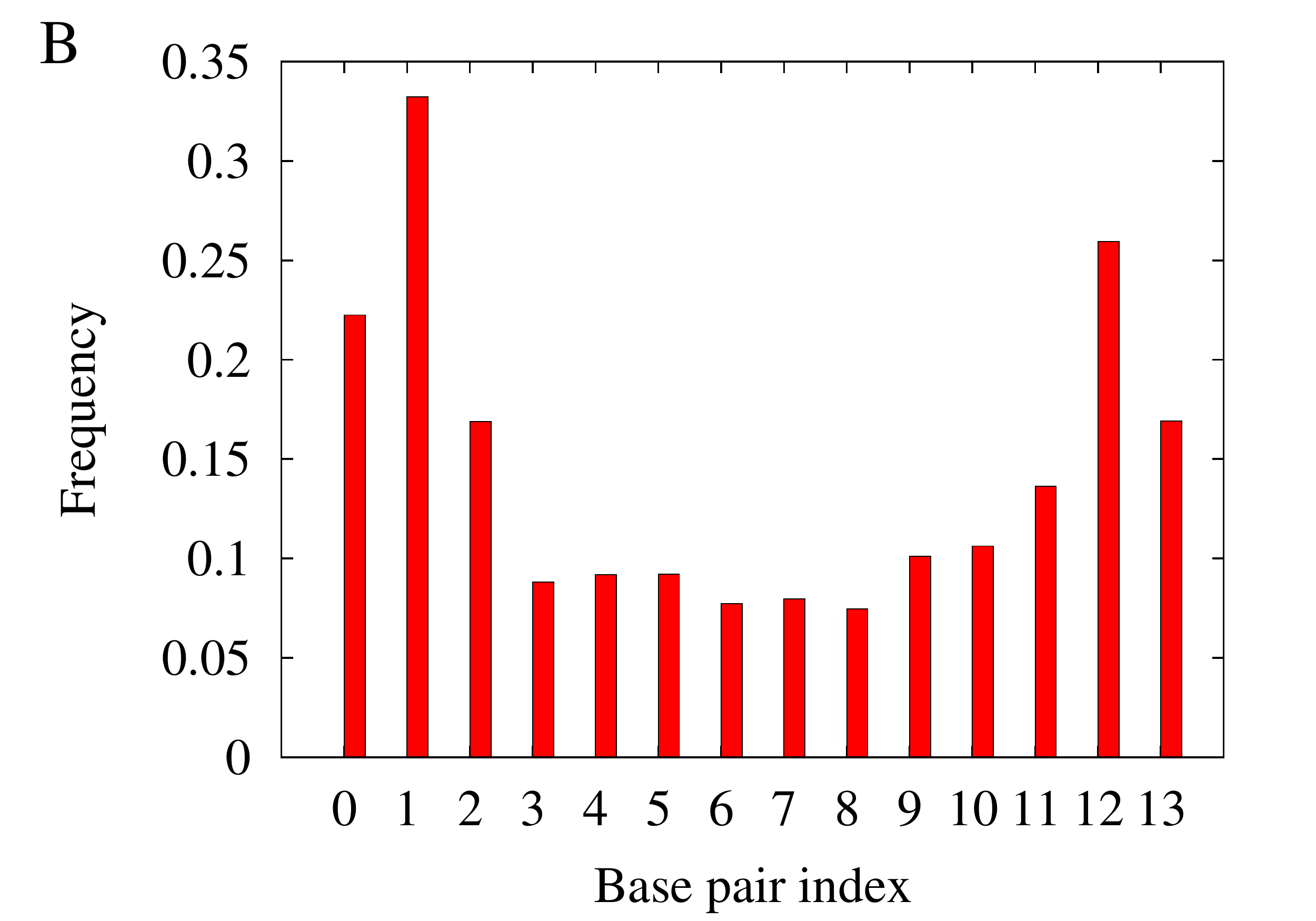}
\includegraphics[angle=-90, width=0.3\textwidth]{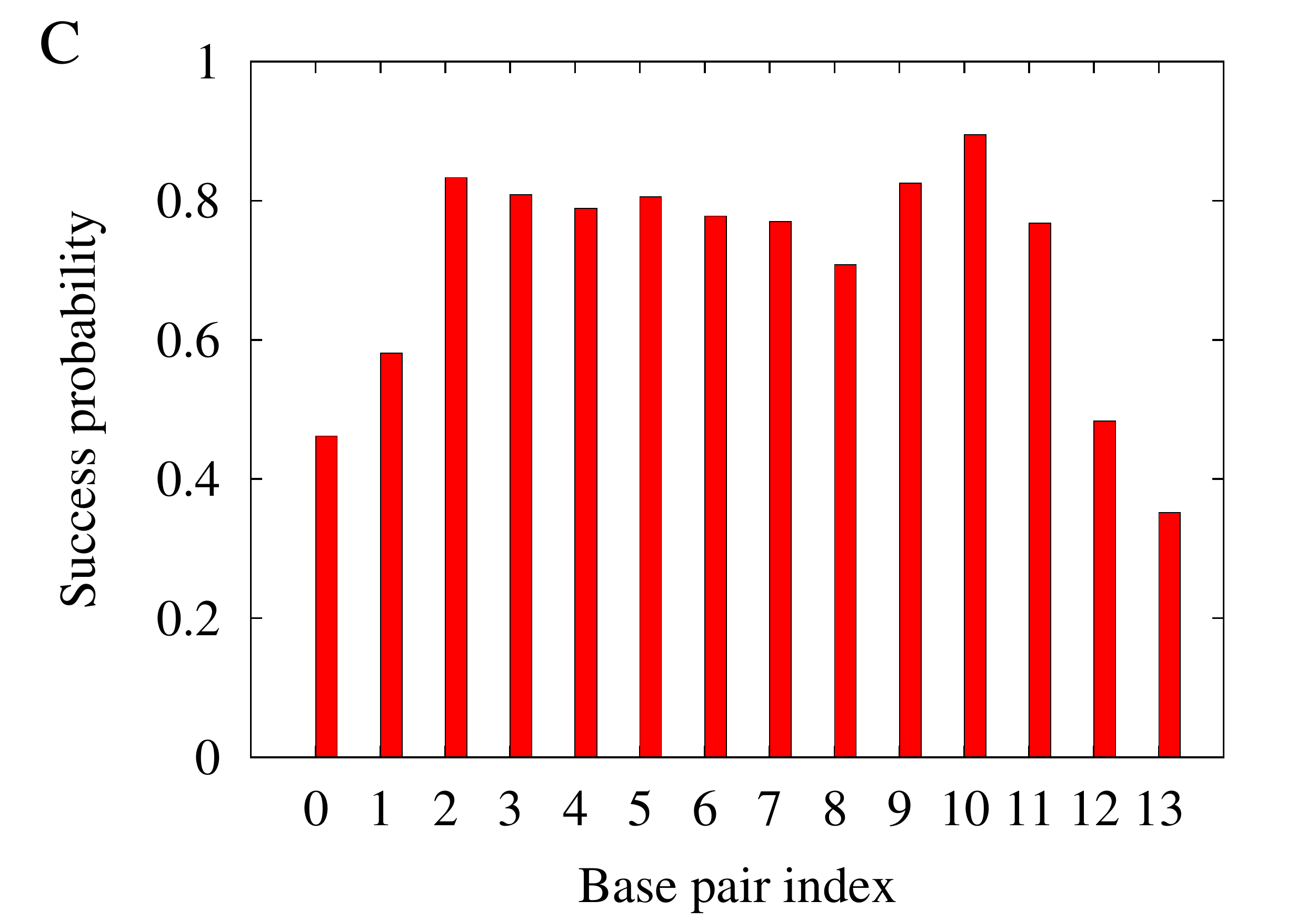}
\end{center}
\caption{\footnotesize{Binding kinetics for non-repetitive sequences. A) ``Arrhenius plot" of the dependence of hybridization rate on temperature, with $k_{\rm on}(T) =  7.67 \times 10^{4}$\,s$^{-1}$ being the rate at 300\,K. The Arrhenius model with a constant activation enthalpy predicts a straight line on such a plot. B) Frequency with which a certain base pair appears in states drawn from the penultimate interface of FFS simulations, for a sequence with native-only reactions at 300\,K. Base pairs with indexes 0 and 14 are at either end of the duplex. Probability of successful duplex formation given the formation of a base pair at the penultimate FFS interface, as a function of base pair index.} \label{fig_flux_data}}
\end{figure*}

The results of hybridization simulations for non-repetitive sequences are given in Table \ref{results-non-repetitive}. We note that the absolute rate at $T=300$\,K ($7.67 \times 10^{4}$\,s$^{-1}$) would, given the concentrations used in the simulations, translate into a bimolecular association rate of $k_{\rm bi} = 7.71\times 10^8$\,M$^{-1}$\,s$^{-1}$. This value is approximately 100 times larger than typical experimental measurements.

As discussed in Appendix \ref{Simulation Methods}, we expect coarse-grained models to provide faster dynamics than real systems. To speed up simulations we used a diffusion coefficient that is 16 times higher than the experimentally measured one, which accounts for much of the difference.  We argued there that our predictions are most reliable when taken as {\it relative} rates.  A graph of $\ln k_{\rm on}$ against $1/T$, with $k_{\rm on}$ the measured association rate, is plotted in Fig.\,\ref{fig_flux_data}\,A. A pure Arrhenius model with a single, well-defined transition state would give a straight line in such a plot. Our result is evidence of the complexity of the ensemble of transition pathways. The frequency of initial contacts (as a function of location within the strand) and the probability of successful duplex formation given those contacts are plotted in Figs.\,\ref{fig_flux_data}\,B and \ref{fig_flux_data}\,C. These histograms show that initial contacts are more likely to occur at the ends of the strands, but more likely to succeed if they occur in the centre. However, all initial attachments occur in the transition pathway ensemble with a reasonable frequency.

\subsection{Characterisation of the ensemble of transition pathways, the equilibrium duplex ensemble and the equilibrium single-stranded ensemble}
\label{results:eqm}
\begin{figure}[h]
\begin{center}
\includegraphics[angle=-90, width=0.4\textwidth]{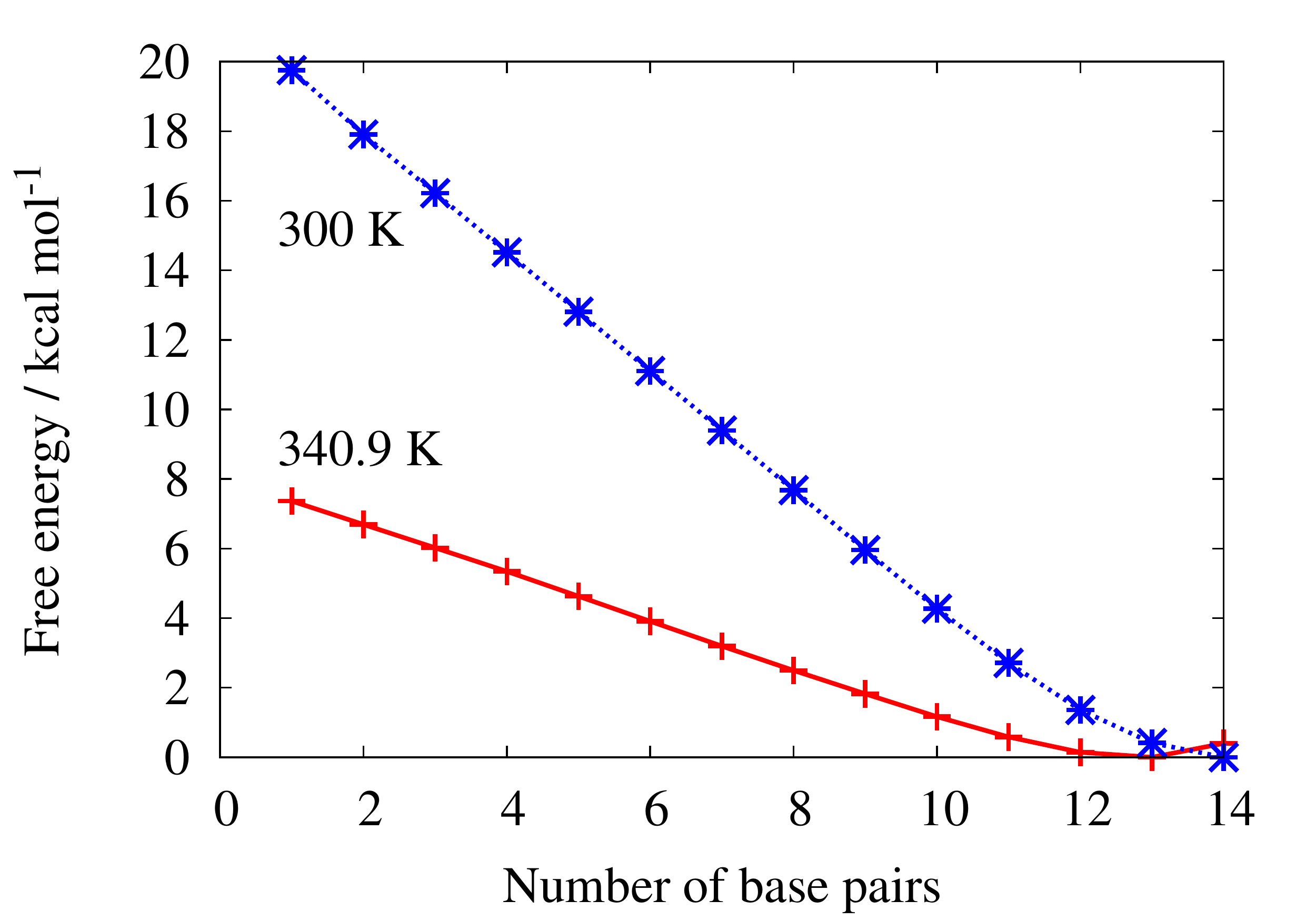}\vspace{-0.1in}
\end{center}
\caption{\footnotesize{Free-energy profile of the duplex state for a 14-base pair duplex with interactions exclusively between native base pairs, measured at 300\,K and 340.9\,K. A base pair is present if its energy is below $-0.596$\,kcal\,mol$^{-1}$. Errors on the points are of order 0.1\,kcal\,mol$^{-1}$. The arbitrary offsets of the free energies are chosen so that the minimum of both curves has a value of zero. These simulations did not sample the unbound state, and hence are concentration independent. We do not show the single-stranded state (0 base pairs) because this is concentration dependent, in contrast to the rest of the free-energy profile. For examples with this state included see e.g. Refs \onlinecite{Ouldridge2011,Ouldridge_thesis,Sulc2012}}\label{fig_FEP}}
\end{figure}

Equilibrium probabilities $P(n)$ of $n$ base pairs with an energy of $E<-0.596$\,kcal\,mol$^{-1}$ being present in the duplex ensemble (when only native base pairs are permitted) at 300\,K and 340.9\,K  are plotted plotted in Fig.\,\ref{fig_FEP}  as a free energy $F(n)/RT = - \ln (P(n))$. 
Fig. \ref{fig_FEP} shows explicitly that adding a new base pair results in a substantial gain in free energy, even at 340.9K. Only taking into account this secondary-structure analysis of the thermodynamic driving force implies that metastable states with one or two base-pairs formed should have a high probability of ending up in the duplex state.

Despite this argument based on equilibrium free energies of secondary structure, analysis of the FFS simulations of association indicate that states with initial contacts are surprisingly likely to fail to form a full duplex. The penultimate FFS interface  in the simulations of association corresponds to configurations in which two base pairs are present with energy significantly more negative than $E<-0.596$\,kcal\,mol$^{-1}$. As can be seen from Table \ref{results-non-repetitive}, simulations launched from this interface have a 33\% chance of successfully zippering to form the full duplex at 300\,K, dropping to 7.8\% at 340.9\,K. Even in the absence of non-native bonds, the success rate is only 65\% for configurations loaded at this interface at 300\,K. As systems at this interface can either proceed to full duplex formation or separate with roughly equal probability at 300\,K, configurations stored at this interface are approximately representative of the `transition ensemble' of the system. 

\begin{table}[!]
\begin{center}
\begin{tabular}{c c  c }
Ensemble & Interstrand   & Av. base pair  \\
& enthalpy/kcal\,mol$^{-1}$ & separation /nm\\
\hline
\\
Binding (kinetic) & -7.04 & 2.84 \\
Duplex (equilibrium) &-10.2 & 2.10 \\
\hline
\end{tabular}
\caption{\footnotesize Properties of ensembles observed during binding and at equilibrium. The binding ensemble consists of states obtained from kinetic simulations at the penultimate FFS interface  (the last one before duplex formation), the equilibrium ensemble are states that satisfy the same base-pairing criteria, but drawn from an equilibrium sampling of the duplex state.  \label{results-final-interface}}
\end{center}
\end{table}

\begin{table}[!]
\begin{center}
\begin{tabular}{c  c   c}
Register &\multicolumn{2}{c}{Flux / s$^{-1}$}\\
& 300\,K & 340.9\,K \\
\hline
Correct bonding\\
0 & $(5.61\pm 0.49) \times 10^4$ & $1.15 \times 10^4$\\
misaligned bonding \\
2 & $(4.99\pm 0.39) \times 10^4$ & $1.51 \times 10^4$\\
-2 & $(3.68 \pm 0.38) \times 10^4$ & $1.24 \times 10^4$\\
4 & $(3.94 \pm 0.44) \times 10^4$ & $8.32 \times 10^3$\\
-4 & $(3.96 \pm 0.41) \times 10^4$ & $1.35 \times 10^4$\\
6 & $(3.58 \pm 0.45) \times 10^4$ & $6.41 \times 10^3$\\
-6 & $(3.17 \pm 0.40)\times 10^4$ & $8.64 \times 10^3$\\
8 & $(3.02 \pm 0.32 )\times 10^4$ & $\sim$\\
-8 & $(2.68 \pm 0.35)\times 10^4$ & $\sim$\\
10 & $(3.27 \pm 0.46)\times 10^4$ & $\sim$\\
-10 & $(2.34 \pm 0.23)\times 10^4$ & $\sim$\\
Pseudoknot bonding \\
-4,8& 173  & $\sim$ \\
-6,6&  801  & $\sim$\\
-6,8& 308  & $\sim$\\
-8,4& 14.6  & $\sim$\\
-8,6&  7.29  & $\sim$\\
-8,8&   & $\sim$ \\
\hline
\end{tabular}
\caption{\footnotesize Total flux from unbound to bound states of various alignments for a repetitive sequence. We interpret these fluxes as rates of instantaneous reactions. Pseudoknots rarely form before any complete register: due to this rarity, relative errors for the pseudoknotted results are approximately 100\%. \label{results-repetitive}}
\end{center}
\end{table}

\begin{table*}[!]
\begin{center}
\begin{tabular}{c | c c c c c c c c c c  }
& \multicolumn{10}{c} {Initial register} \\
 result&  2 & -2 & 4 & -4 & 6 & -6 & 8 & -8 & 10 & -10 \\
\hline
0  & \hilighty {$1.85 \times 10^4$} & \hilighty{$2.68 \times 10^4$} &$1.55 \times 10^3$ & 298 & 128 & 41.4 & $1.04 \times 10^4$ & $5.04 \times 10^3$& \hilightg{$4.74 \times 10^5$}  & \hilightg{$5.30 \times 10^5$}\\
2  & $\sim$ & & \hilighty{$4.95 \times 10^4$} & & 867 &    &$1.15 \times 10^3$ & \hilightg{$9.51 \times 10^4$} & \hilightg{$2.59 \times 10^5$} & \hilightg{$3.05 \times 10^5$} \\
-2 &34.0 & $\sim$& &\hilighty{$4.54 \times 10^4$} & 30.9 & 957 & \hilightg{$1.01 \times 10^5$} &$3.95 \times 10^3$ & \hilightg{$3.33 \times 10^5$} & \hilightg{$3.23 \times 10^5$}\\
4 & 299 & & $\sim$ & & \hilighty{$8.23 \times 10^4$} & $1.43 \times 10^3$ & $2.85 \times 10^3$ & $2.82 \times 10^3$ &$2.88 \times 10^4$ & \hilightg{$4.81 \times 10^5$}\\
-4 &  & 2.16 & &$\sim$  &  $2.96 \times 10^3$& \hilighty{$8.89 \times 10^4$}& $1.91 \times 10^3$ & $4.42 \times 10^3$ & \hilightg{$2.81 \times 10^5$} & $3.60 \times 10^4$ \\
6  & & & 45.9 & & $\sim$ & 5.46 & \hilightg{$1.31 \times 10^5$} & $1.38 \times 10^3$ & $1.55 \times 10^4$ & \hilightg{$1.76 \times 10^5$}\\
-6  & & & &  212 & &$\sim$ &  &\hilightg{$1.55 \times 10^5$} & \hilightg{$2.69 \times 10^5$} & $2.60 \times 10^4$\\
8 &  & & & & 93.7 & & $\sim$ & & \hilightg{$5.49\times 10^5$} & $1.00 \times 10^5$\\
-8 & & & & & & 586 & & $\sim$& $8.10 \times 10^4$ & \hilightg{$6.54 \times 10^5$}\\
10  & & & & & & & 409 & 183 & $\sim$ & $1.99 \times 10^4$\\
-10  & & & & & & & & 211 & $9.98 \times 10^3$ & $\sim$\\
-4,8 & & & & & & & \hilightg{$9.91 \times 10^4$} & &$2.40 \times 10^4$  & $1.87 \times 10^3$\\
-6,6 & & & & &$1.72 \times 10^3$ &  $1.96 \times 10^4$& 462& $1.60 \times 10^3$&  & $5.92 \times 10^3$\\
-6,8 & & & & & & $1.28 \times 10^3$ & \hilighty{$1.52 \times 10^5$} & & $2.89 \times 10^4$ & $2.85 \times 10^3$\\
-8,4 & & & & & 11.7& 84.3 & & \hilightg{$5.58 \times 10^4$} & $1.21 \times 10^4$ & {$1.52 \times 10^4$}\\
-8,6 & & & & & $1.64 \times 10^4$& & & \hilighty{$1.80 \times 10^5$} & & $1.04 \times 10^4$\\
-8,8 & & & & & & & \hilightg{$9.84\times 10^4$} & \hilightg{$8.81 \times 10^4$} & $1.42 \times 10^4$ & $7.65 \times 10^3$\\
melt & $\sim$ & $\sim$ & $\sim$ & $\sim$ & $\sim$ & $\sim$ &$8.92 \times 10^3$  & $9.04 \times 10^3$ & \hilighty{$1.47 \times 10^6$} & \hilighty{$1.71 \times 10^6$}\\
\hline
\end{tabular}
\caption{\footnotesize Total flux from misaligned states  to other (meta)stable states for the repetitive sequence at 300\,K. Blank spaces indicates events that could potentially have occurred during simulations, but were not observed. `$\sim$' indicates transitions that were not sampled. The most common transition for each initial state is highlighted in yellow, other reasonably frequent results are highlighted in green. Standard errors on the estimates for internal displacement are on the order of 15\% for the most common results, rising to 100\% for less frequently observed traditions. Standard errors on the estimates of melting are in the range 10 -- 25\%. \label{results-repetitive_2}}
\end{center}
\end{table*}

\begin{table}[!]
\begin{center}
\begin{tabular}{c | c  }
Register &Flux / s$^{-1}$\\
\hline
Correct bonding\\
0 & $(1.15 \pm 0.19) \times 10^4$\\
misaligned bonding \\
2 & $(1.51 \pm 0.35) \times 10^4$\\
-2 & $(1.24 \pm 0.24) \times 10^4$\\
4 & $(8.32 \pm 1.4) \times 10^3$\\
-4 & $(1.35 \pm 0.33) \times 10^4$\\
6 & $(6.41 \pm 2.8) \times 10^3$\\
-6  & $(8.64 \pm 1.5) \times 10^3$\\
\hline
\end{tabular}
\caption{\footnotesize Total flux from unbound to various misaligned states for the repetitive sequence at 340.9\,K. We interpret these fluxes as rates of instantaneous reactions. \label{results-repetitive_highT}}
\end{center}
\end{table}

\begin{table}[!]
\begin{center}
\begin{tabular}{c | c c c    }
 &  2  & 4  & 6  \\
 \\
\hline
0  &  \hilighty{$1.66\times 10^5$} & $2.37 \times 10^4$ & $1.37 \times 10^4$\\
2 & $\sim$& \hilightg{$4.54 \times 10^5$} & $1.27 \times 10^5$\\
-2 &$2.39 \times 10^3$ & $4.54 \times 10^3$ & $1.21 \times 10^4$\\
4 & $4.22\times 10^4$& $\sim$ & \hilightg{$7.73 \times 10^5$}\\
-4 & 763 & 179 & $1.27 \times 10^5$\\
6 & & $6.24 \times 10^4$ & $\sim$ \\
-6 & 84.6 & $1.42\times 10^3$ & $5.03\times 10^4$\\
melt  & \hilightg{$6.39\times 10^4$}  & \hilighty{$4.40\times 10^5$} & \hilighty{$3.64 \times 10^6$} \\
\hline
\end{tabular}
\caption{\footnotesize Total flux from misaligned states  to other (meta)stable states for the repetitive sequence at 340.9\,K. Blank spaces indicates events that could potentially have occurred during simulations, but were not observed. `$\sim$' indicates transitions that were not sampled. The most common transition for each initial state is highlighted in yellow, other reasonably frequent results are highlighted in green. Standard errors on the estimates for internal displacement are on the order of 10\% for the most common results, rising to 100\% for less frequently observed traditions. Standard errors on the estimates of melting are approximately 5\%. \label{results-repetitive_highT_2}}
\end{center}
\end{table}

In the main text, we argue that these transition states are not representative of equilibrium states with the same degree of base pairing. To establish this, we compare states from the penultimate interface of FFS simulations of association in which only native base-pairs were permitted with states obtained from the equilibrium ensemble that satisfy the same base-pairing criteria. For the two ensembles, the average separation of native base-pair contacts and the average overall interstrand enthalpy were measured. The results are given in Table \ref{results-final-interface}.

The difference in interstrand enthalpies between the two ensembles is primarily due to stronger stabilizing cross-stacking interactions in the equilibrium configurations. It is not this difference in enthalpy itself, however, that explains our results. In the FFS simulations of dissociation of register $-10$ at 300\,K, for simulations launched from the penultimate interface prior to separation ($\lambda_1^2$) full dissociation was observed in  2237 trajectories and reformation of the full duplex in 5173, despite an average interstrand enthalpy of only  $-1.28$\,kcal\,mol$^{-1}$ at this interface. Note that many trajectories launched from $\lambda^2_1$ reached alternative metastable states, rather than dissociating -- in the figures given above we only consider trajectories launched from configurations at $\lambda^2_1$ in which no other register of bonding is present, explaining the difference between the numbers quoted and those in Table \ref{FFS-melt-data}. Rather, the weaker interactions and greater distance between hydrogen-bonding sites in the kinetic ensemble indicate that the geometry of the two strands is not generally conducive to full hybridization, and not reflective of states with comparable enthalpy in the bound ensemble. As a result, there is a reasonable probability of strands dissociating even after making initial contacts with significant interstrand interactions, and therefore the process of duplex formation has a non-negligible negative activation enthalpy. 

We have also noted a competing contribution to the activation enthalpy by comparing the intrastrand interactions in the equilibrated single-stranded state with those in the ensemble of states from the penultimate FFS interface. For the unbound ensemble, we obtain an average of $-133$\,kcal\,mol$^{-1}$, compared to $-129$\,kcal\,mol$^{-1}$ from the states at the penultimate interface of FFS simulations. In the absence of intrastrand base-pairing, this difference is attributable to less effective stacking of the individual strands in the hybridization ensemble. Disrupting stacking makes it easier for the strands to be in contact  without being fully bound, and also optimal stacking configurations are not consistent with duplex geometry.\cite{Ouldridge2011} The difference in intrastrand enthalpies contributes to the overall activation enthalpy of binding, tending to make it less negative. However, for our model, the effect of disrupting stacking is smaller contribution than other effects that favour a negative activation enthalpy.

\begin{table*}[t]
\begin{tabular}{c c  c c}
Sequence &Type & Relative binding rate & Bimolecular rate constants \\
& & & from Ref.\,\onlinecite{Zhang_disp_2009} / M$^{-1}$s$^{-1}$ \\
\hline
\multicolumn{3}{c} {misaligned bonds allowed} \\
$5^\prime$-CCCGCCGC-$3^\prime$ & G-C-rich & 1 & $6\times 10^6$ \\
$5^\prime$-TCTCCATG-$3^\prime$ & average-strength & $0.28 \pm 0.05$ & $3\times 10^6$ \\
$5^\prime$-ATTTATTA-$3^\prime$  & A-T-rich & $0.14 \pm 0.02$  & $4\times 10^5$\\
\multicolumn{3}{c} {misaligned bonds not allowed}\\
$5^\prime$-CCCGCCGC-$3^\prime$ & G-C-rich & $0.32 \pm 0.06$\\
$5^\prime$-ATTTATTA-$3^\prime$ & A-T-rich & $0.10 \pm 0.02$ \\
\hline
\end{tabular}
\caption{Binding rates of 8-base strands of different sequences at 298.15\,K, relative to the G-C-rich case. Reaction rates were measured for the  strands shown and their complements. Reaction constants are taken from the long-toehold limit of the fits in Ref.\,\onlinecite{Zhang_disp_2009}\label{results-seq-dependent}}
\end{table*}

\subsection{Hybridization of repetitive sequences}
\label{results-rep}
The results of the FFS simulations of the initial hybridization of repetitive sequences at 300\,K are given in Table \ref{results-repetitive}. The probability of formation of each register $r_1$ is approximately proportional the number  of bonds available, $14 -|r_1|$. The results of FFS simulations of rearrangement and dissociation at 300\,K are summarised in Table \ref{results-repetitive_2}. Equivalent results for simulations at 340.9\,K are given in Tables \ref{results-repetitive_highT} and \ref{results-repetitive_highT_2}. 

As is evident, initial misaligned structures with more than four base pairs tend to rearrange into structures with a greater degree of base-pairing at 300\,K. Registers $r_1=\pm 10$, with only four base pairs, have a similar probability of forming a more strongly-bound duplex and dissociating. We were unable to reliably simulate the relaxation of the relatively stable pseudoknot structures $\{r_1,r_2\} = \{-4,8\}$, $\{-6,6\}$, $\{-6,8\}$, $\{-8,4\}$ and $\{-8,6\}$. However, given that each register present in these metastable pseudoknots can form at least six base pairs, it seems likely that these structures would eventually relax to the fully-formed duplex. We emphasize that unlike the six pseudoknots listed above, most pseidoknots relax to a single register reasonably quickly.

To estimate the rate of formation of the fully-formed duplex, we therefore sum over the rate of formation of all structures from the initial simulations, with the exception of registers $r_1 = \pm 10$. In these cases we take the fraction of structures that rearrange into another structure with a higher degree of base-pairing as the fraction that eventually form a full duplex. The result thus obtained is $k_{\rm on} = 3.8 \times 10^5$\,s$^{-1}$, approximately five times larger than the result for non-repetitive sequences given in Table \ref{results-non-repetitive}. As justified in Appendix \ref{subtleties}, this analysis ignores the time spent in the metastable intermediates.

At 340.9\,K, dissociation is a non-negligible pathway even for the most stable misbonds, $r_1 = \pm 2$. To analyze this case, we considered only the two most likely routes out of each metastable state, those highlighted in green and yellow in Table \ref{results-repetitive_highT_2}. Calculation of the overall transition rate into the $r_1=0$ state is then a relatively simple problem, yielding $k_{\rm on} = 4.0 \times 10^4$\,s$^{-1}$, almost 10 times smaller than for the same sequences at 300\,K (this calculation assumes the negative registers behave identically).

\subsection{Sequence-dependence of association rate}
\label{results:seq-dep}
The  relative association rates of eight base-pair duplexes with varying sequence, obtained using the protocols outlined in Appendix \ref{protocols:seq-dep},  are presented in Table \ref{results-seq-dependent}. Also shown are the rates of displacement fitted by Zhang and Winfree\cite{Zhang_disp_2009} for G-C-rich, A-T-rich and average-strength toeholds, in the limit of long toehold lengths (we take our eight-base sequences from this source).  These rates are assumed to reflect the association rates of the toeholds themselves.\cite{Zhang_disp_2009}

\section{Detailed comparison of oxDNA with 3SPN.1}
\label{3SPN vs oxDNA}
Here we discuss the differences between our results and those for 3SPN.1, an alternative model of DNA. This discussion is needed because 3SPN.1 has also been used to study hybridisation,\cite{Sambriski2008,Sambriski2009_PNAS,Schmitt2011, Hoefert2011,Schmitt2013} finding some similar results (such as transitions being complex) but, importantly, finding significantly different pathways towards hybridization. There are several major differences between oxDNA and 3SPN.1 that are relevant to this analysis.
\begin{itemize}
\item Single-stranded DNA in 3SPN.1 consists of unphysically stiff helices,\cite{Sambriski2009_biophys} whereas single strands in oxDNA can unstack and hence are more flexible, with a greater degree of conformational freedom. The importance of treating the extra flexibility of ssDNA relative to duplexes is evident in the formation of single-stranded hairpins\cite{SantaLucia2004} and in the force-extension properties of ssDNA,\cite{Smith1996,
Dessinges2002} both of which are accurately reproduced by oxDNA.\cite{Ouldridge2011,Romano_overstretch_2012}
\item The base-pairing interaction in oxDNA is strongly modulated by orientation of the nucleotides \cite{Ouldridge2011}, meaning that the edges of bases must point at each other to form bonds. This reflects the strongly directional nature of hydrogen bonding.  3SPN.1 has several beads for each nucleotide, but all interactions between beads are isotropic. Thus  bonding can occur in configurations in which the bases are close to each other, but not in a realistic orientation for hydrogen-bonding. As discussed by Florescu and Joyeux \cite{Florescu2011}, these isotropic interactions can even lead to unphysical stable states for poly(dA)-poly(dT) in which each nucleotide is bound to two others (although Florescu and Joyeux studied an earlier version of the model, 3SPN.0,\cite{Knotts2007} the hydrogen-bonding geometry is unchanged in 3SPN.1).
\item 3SPN.1 contains an attractive interaction between sugar sites that was introduced to mediate the hybridization reaction.\cite{Sambriski2009_biophys} This attraction provides a stabilizing contribution to the system when the single strands are in close proximity to each other, but not bound with hydrogen bonds. Sambriski {\it et al.}\cite{Sambriski2009_biophys} justified this term by referring to the tendency of DNA duplexes to condense in the presence of multivalent ions, but its role in a model parameterized for monovalent ions is unclear. 
\end{itemize}

Next we discuss how these differences play out for the dynamics of hybridisation. Perhaps the most important geometric difference between the two models is the fact that the single strands in 3SPN.1 are stiff and helical.   We show that zippering in oxDNA occurs because the single-strands are relatively flexible: double helices form in stages as bases stack onto the end of the growing duplex. The stiffness of the duplex itself is an emergent property, rather than being imprinted at the level of the single strands. By contrast, in 3SPN.1, hybridisation occurs through the association of two fairly stiff helices, for which the most natural pathway is probably the winding referred to in the detailed study by  Schmitt and Knotts.\cite{Schmitt2011}

Another way the flexibility of the strands plays an important role  involves the mechanism of internal rearrangement.
The intermediates of internal displacement, involving bulged or pseudoknotted states, require significant flexibility in the single strands, and hence these processes will be suppressed by the stiff single strands in 3SPN.1.

 Instead of using internal displacement, repetitive strands in 3SPN.1  can slither past each other\cite{Sambriski2008,Sambriski2009_PNAS,Hoefert2011} in a mechanism `devoid of significant energy barriers'.\cite{Hoefert2011}  This ability to `slither' suggests that a similar sliding mechanism may also explain how initially misaligned non-repetitive duplexes relax to the native state.\cite{Schmitt2011,Schmitt2013} Slithering is not observed in oxDNA.  In order to undergo slithering, the strands must slide relative to each other along the duplex axis.  Performing such an operation with oxDNA would be extremely costly: the system would have to move through an intermediate state in which all base pairs were broken but the strands were still held in a double helical orientation, wrapped around each other. Thus reaching the intermediate state involves an enormous enthalpic cost, with little entropic gain to compensate. By contrast, in 3SPN.1 this process, is `devoid of significant energy barriers' for repetitive sequences with a repeat unit of two bases.\cite{Hoefert2011} Several factors contribute to this difference. Firstly, the isotropic nature of interactions means that hydrogen-bonding need not be fully disrupted during slithering. Secondly, the attraction between sugar sites stabilizes a state in which the two strands are wrapped round each other, but not base-paired. Finally, the fact that 3SPN.1 helices are so stiff means that the conformational freedom of single strands is significantly reduced. Therefore the fact that they must remain helical during the slithering process incurs a relatively  smaller entropic penalty than in oxDNA, meaning that it is a viable alternative to dissociating. We note that, although internal displacement via inchworm and pseudoknot intermediates can occur in oxDNA, both processes  nevertheless involve significant free-energy barriers associated  with initiating the displacement.

Any coarse-grained DNA model makes compromises between accuracy and tractability.   In fact such models will never simultaneously reproduce all the properties of DNA, a general attribute of effective coarse-grained systems sometimes called ``representability problems''.\cite{Louis2002} 
OxDNA was specifically designed in order to reproduce hybridization thermodynamics as well as the mechanical properties of both single and double strands.   We argue here that capturing the strand flexibility as well as the orientational dependence of the effective potentials  is crucial  if one wants to reproducing the gross features of the hybridization kinetics we focus on in this paper.  Our success at quantitatively reproducing relative rates measured for strand displacement systems~\cite{Zhang_disp_2009} gives us confidence in our predictions of similar physical phenomena in hybridisation. 

3SPN.1 has  some advantages over oxDNA. For example, 3SPN.1 explicitly represents the asymmetric grooves in DNA, allowing structural properties that are sensitive to this feature to be modelled. Electrostatic screening effects are also explicitly included, allowing 3SPN.1 to capture the effects of changing salt concentrations, whereas oxDNA is limited to one salt concentration. Encouragingly, both models show that hybridisation can proceed through complex pathways. Nevertheless,  we conclude that oxDNA's representation of hybridization, involving nucleation and zippering of flexible strands to form stiff helices and the possibility of internal displacement, is more likely to represent true features of real DNA.  Of course at the end of the day, the true arbiter of all these predictions will be experiment, and it is likely that slithering and internal displacement will give distinguishable predictions as features such as the repeat length of a repetitive sequence are changed.

\end{document}